# Quantization of Sinai's Billiard - A Scattering Approach

Holger Schanz
*Institut für Theoretische Physik*
*Humboldt - Universität*
*Invalidenstr. 110*
*10 099 Berlin, Germany*

and

Uzy Smilansky
*Department of Physics of Complex Systems,*
*The Weizmann Institute of Science*
*Rehovot, 76 100 Israel*

**Abstract**

We obtained the spectrum of the Sinai billiard as the zeroes of a secular equation, which is based on the scattering matrix of a related scattering problem. We show that this quantization method provides an efficient numerical scheme, and its implementation for the present case gives a few thousands of levels without encountering any serious difficulty. We use the numerical data to check some approximations which are essential for the derivation of a semiclassical quantization method based also on this scattering approach.

## 1 Introduction

The Sinai Billiard (SB) has become a benchmark for testing ideas and methods in "quantum chaos". Bohigas and his coworkers [1] calculated the lowest few hundreds of levels by using the boundary integral method [2], and showed that random matrix theory accounts very well for the numerical level statistics. Berry [3] used the KKR method [4] as a starting point for his investigation of the semiclassical limit. More recently, the scattering approach to quantization [5] was demonstrated by applying it for the quantization of the SB. Last, but not least, this system was the first to be investigated experimentally by measuring $\mu$ - wave reflection from a billiard shaped cavity [7].

In the present paper we would like to present yet another study of the SB, aiming this time to check and demonstrate the scattering approach to quantization in greater detail and depth. The scattering approach to quantization is based on the observation [5] that *every* billiard interior problem can be viewed upon as a scattering (exterior) problem, and the spectrum can be uniquely deduced from the knowledge of the corresponding scattering operator. As a matter of fact, one can show that the spectrum of the billiard coincides with the set of energies for which an eigenvalue of the scattering operator takes the value 1.

The scattering system associated with a given billiard may be constructed in various ways. One may consider the billiard boundary as an obstacle in a scattering (exterior) problem [5]. Dirichlet (or Neumann) boundary conditions define uniquely the scattering operator and through Pillet's theorem [6], the interior energies are given as the energies at which an eigenvalue of the scattering operator obtains the value 1. The other approach, which will be pursued and investigated here, is based upon attaching appropriate "channels" to the billiard and the scattering operator coincides with the transmission or reflection matrix, corresponding to this quasi 1-d scattering problem [5]. As a matter of fact, the exact quantization involves the extension of the concept of the scattering matrix to the space of "closed" or "evanescent" modes. When this is done one can prove (see chapter (2) that the spectrum of the interior problem coincides with the energies, for which an eigenvalue of the extended scattering operator takes the value 1.

The scattering approach has a few advantages which will be demonstrated in the present note. One certain advantage is the numerical efficiency of this method. In chapter (3) we shall discuss some numerical results, and use the calculated spectrum and $S$ matrices to demonstrate a few important relations. The scattering approach can be used naturally as a basis for the derivation of the semiclassical limit. To get to this limit, one has to go through two stages. In the first (the "semiquantal" approximation [5]) one neglects all the contributions from evanescent modes. The scattering operator is reduced to a finite matrix and the secular equation for the interior problem becomes:

$$Z_{sq}(E) = det(I - S(E)) = 0 \tag{1}$$

Here, $S(E)$ is a $\Lambda \times \Lambda$ unitary matrix, whose dimension $\Lambda$ is in general a step function of the energy, increasing by one at each threshold energy when a "closed" mode starts to conduct. The accuracy of the semiquantal approximation is expected to deteriorate at the vicinity of threshold energies [5], in a way which will be discussed in chapter (4). The semiclassical approximation is introduced by evaluating (1) semiclassically. One can show (see chapter (4)) that (1) can be expressed in terms of $Tr(S^n)$, $1 \leq n \leq \Lambda$, and the traces $Tr(S^n)$ can be written in terms of periodic orbits of the billiard! One derives in this fashion a semiclassical secular equation based on a finite number of periodic orbits, without having to worry about divergent series and the proliferation of the number of periodic orbits.

We shall end this paper with a summary, in which we shall list some interesting open problems and possible directions where the scattering approach to quantization might compete advantageously with other methods. We shall also discuss the relationships between the scattering approach and the semiclassical methods of Bogomolny [8] and others.

# 2 Exact Quantization

We consider a quarter of Sinai's billiard, enclosed by the four sides of the square with diagonal $\{0,0\}$ - $\{L,L\}$ and a quarter circle of radius $R \leq L$ centered at the origin (cmp. Fig. 1). We are looking for a wave function $\Psi^k(x,y)$ vanishing at the boundary and satisfying

$$(\Delta + k^2)\Psi^k = 0 \tag{2}$$

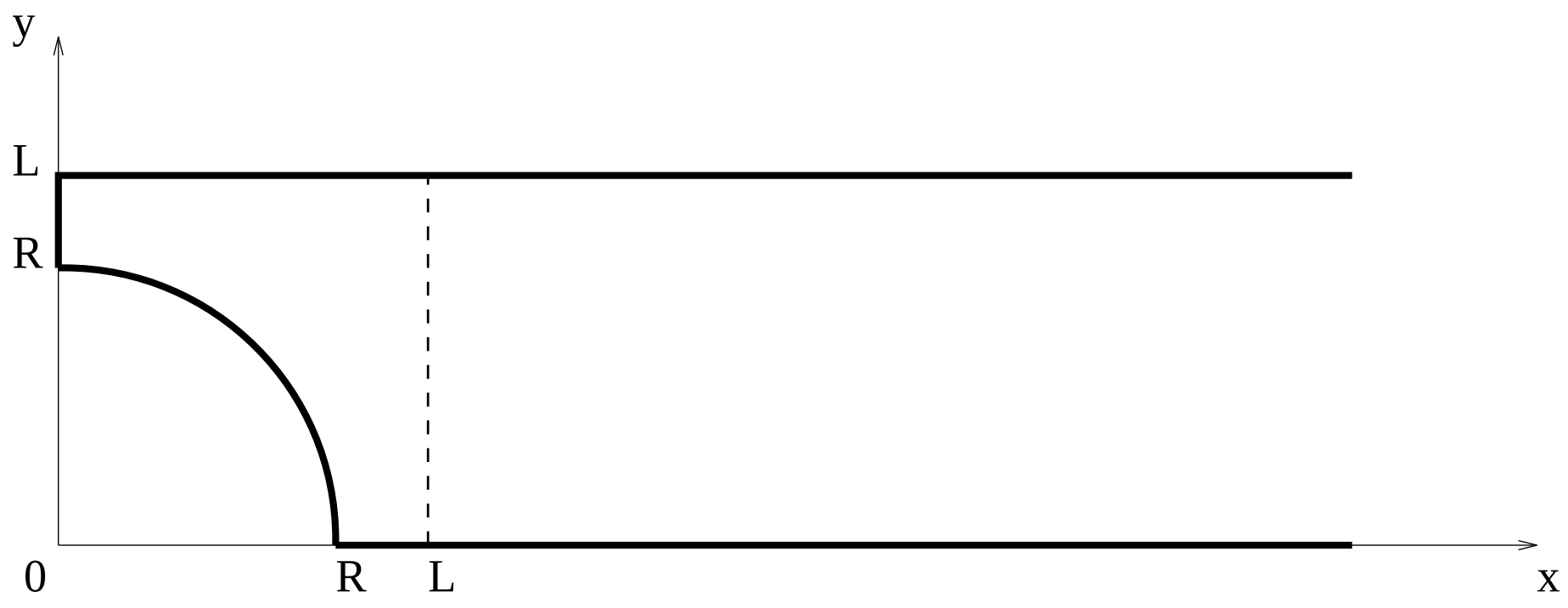

Figure 1: *The Sinai billiard and the attached channel.*

In the region $x \geq R$ it can be decomposed into normal modes

$$\phi_n(y) = \sqrt{\frac{2}{L}} \sin \frac{n\pi y}{L}. \tag{3}$$

The wave function will then take the form

$$\Psi^k(x,y) = \sum_{n=1}^{\infty} b_n \left\{ \frac{e^{-ik_n x}}{\sqrt{k_n}} \phi_n(y) - \sum_{m=1}^{\infty} S^L_{mn} \frac{e^{ik_m x}}{\sqrt{k_m}} \phi_m(y) \right\}, \tag{4}$$

where $b_n$ and $S^L_{mn}$ are some constants. The total energy splits into two parts

$$k^2 \ = \ k_n^2 + (\frac{n\,\pi}{L})^2 \tag{5}$$

where $k_n$ is the wave number for the motion in x - direction. Depending on n it will be a real or purely imaginary number. For $n \leq \Lambda(k)$ with

$$\Lambda(k) = [kL/\pi] \tag{6}$$

we have $k_n^o = |k_n|$ and the mode is called open (traveling) whereas for a closed (evanescent) mode with $n > \Lambda(k)$ $\quad k_n$ obeys $k_n^c = +i|k_n|$. The number of open channels $\Lambda(k)$ depends on the total energy. The open modes consist of an incoming and an outgoing plane wave, the closed ones contain an exponentially increasing and a decaying contribution. For the sake of a compact formulation the terms incoming and outgoing will also be used for the evanescent modes.

To be an eigenfunction of the closed billiard, the wave function must vanish at $x = L$ and this yields a system of linear equations for the coefficients $b_n$:

$$0 = \sum_n b_n \left\{ \delta_{mn} \frac{e^{-ik_mL}}{\sqrt{k_m}} - S^L_{mn} \frac{e^{ik_mL}}{\sqrt{k_m}} \right\} \quad \forall m \tag{7}$$

It will have a solution if

$$det(\delta_{mn} - e^{2ik_mL} S^L_{mn}) = 0. \tag{8}$$

In order to use this equation for the evaluation of the billiard eigenenergies one must know the $S^L_{mn}$ as a function of $k$. They can be defined as the coefficients of the outgoing modes $m$ in the unique solution $\Psi_n$ with a single normalized incoming mode in channel $n$

$$\Psi_n = \frac{e^{-ik_nx}}{\sqrt{k_n}} \phi_n(y) - \sum_m S^L_{mn} \frac{e^{ik_mx}}{\sqrt{k_m}} \phi_m(y) \tag{9}$$

In the limit $x \to \infty$ and for $n \leq \Lambda(k)$ the only terms that survive in (9) are the open modes. Hence $S^L_{mn},\ n, m < \Lambda(k)$ is the unitary scattering matrix. The unrestricted $S^L_{mn}$ is a generalization of the scattering operator for the scattering from the left billiard wall [21]. Obviously $S^R = diag(e^{2ik_nL})$ is a generalization of the scattering matrix for scattering from the right (straight) billiard wall at $x = L$ in the same sense, and so we can write

$$det(I - S^R S^L) = 0. \tag{10}$$

This is the exact secular equation for the billiard eigenvalues in the form of (1). For a usual scattering matrix unitarity and symmetry are important features and provide e. g. a valuable means for checking the numerical accuracy of the computations. In appendix B we show how these relations can be extended to hold for the generalized $S$ as well.

It is well known [9] that the elements of the unitary $S$ - matrix are related to the matrix elements of the Green function in a mixed coordinate - momentum representation. If we impose outgoing boundary conditions, the Green function $G^+$ takes the asymptotic form:

$$G^+(x,y;x',y') = \frac{i}{2}\sum_{m,m'=1}^{\infty} \frac{\phi_m(y)}{\sqrt{k_m}}\frac{\phi_{m'}(y')}{\sqrt{k_{m'}}}(\delta_{mm'}\, e^{ik_m|x-x'|} - g_{mm'}\, e^{ik_m x + ik_{m'}x'}) \quad (11)$$

where $g_{mm'}$ are constants and depend on an additional boundary or symmetry condition imposed on $G^+$. For example, $g_{mm'} = \delta_{mm'}$ will provide $G_0^+$, the Green function for the "pure" channel closed at $x = 0$ with a straight wall. In our case $g_{mm'}$ are such that the Green function obeys Dirichlet boundary conditions on the left closure of the channel $(x = 0, y > R)$ and $x^2 + y^2 = R^2$. Comparing (11) and (9) and using Green's theorem we obtain:

$$S_{mn} = g_{mn}. \quad (12)$$

This can be also expressed as:

$$\begin{aligned}(I - S^L)_{mm'} \;=\; & \frac{2\sqrt{k_m k_{m'}}}{i} e^{-i(k_m+k_{m'})L} \times \quad (13) \\ & \int_0^L \int_0^L \mathrm{d}y\mathrm{d}y' \phi_m(y)(G^+(L,y;L,y') - G_0^+(L,y;L,y'))\phi_{m'}(y')\end{aligned}$$

Another straight forward application of Green theorem yields:

$$G^+(\vec{r}\,'',\vec{r}\,') - G_0^+(\vec{r}\,'',\vec{r}\,') = \int_\gamma G_0^+(\vec{r}\,'',\vec{r})\frac{\partial}{\partial r}G^+(\vec{r},\vec{r}\,')\mathrm{d}\gamma(\vec{r}) \quad (14)$$

where $\gamma$ is the quarter of the circle $x^2 + y^2 = R^2$, which is traversed in the positive mathematical sense. Equation (14) can be solved formally by successive iterations, resulting in:

$$\begin{aligned}G^+(\vec{r},\vec{r}_0) - G_0^+(\vec{r},\vec{r}_0) \;=\; & \sum_{N=1}^{\infty} \int_\gamma \mathrm{d}\gamma(\vec{r}_1)\cdots\int_\gamma \mathrm{d}\gamma(\vec{r}_N) G_0^+(\vec{r},\vec{r}_N) \\ & \prod_{K=0}^{N-1} \frac{\partial}{\partial r_{N-K}} G^+(\vec{r}_{N-K},\vec{r}_{N-K-1}) \quad (15)\end{aligned}$$

Equations (15) and (14) form a convenient starting point for the semiclassical approximation of $S^L$.

We shall now derive an explicit expression for the $S$ matrix using the methods which were developed in [10]. There, the scattering matrix for an infinite one-dimensional array of non overlapping spherically symmetric scatterers located at $nD\vec{e_y}$ in the two dimensional plane is calculated (cmp. Fig. 2). Our generalized $S$ will be obtained by using the hard disk scattering phase shifts, and by extending the method to evanescent modes followed by a proper desymmetrization to fulfill the boundary conditions at the channel walls. Geometry requires for this purpose $D = 2L$.

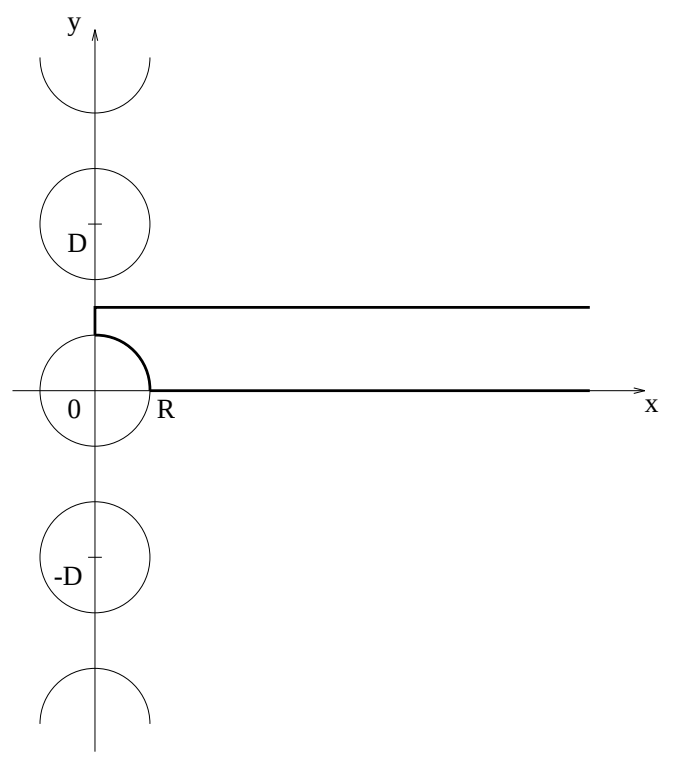

Figure 2: *The array of scatterers and the desymmetrized section of interest.*

Scattering off the infinite periodic array of discs picks certain Bragg manifolds $\alpha$ with discrete y - momenta

$$k_n^y = \alpha + n\frac{2\pi}{D}. \tag{16}$$

The boundary conditions will be fulfilled, if our wave function is antisymmetric with respect to $x = 0$ and both, $y = 0$ and $y = D/2$. Within one Bragg manifold this can only be achieved for $\alpha = 0$. The wave function is then periodic with the period $D$ of the lattice and $\Psi(x, 0) = 0$ will imply $\Psi(x, D/2) = 0$. In the following we restrict ourselves to the case $\alpha = 0$.

A wave function $\Psi_n^l(x, y)$ with an incident plane wave from $x = -\infty$ in channel n is given in [10]-(II.14, 15) in normal mode representation which is valid for $|x| > R$. In our notation it reads

$$\begin{aligned} \Psi_n^l &= \tfrac{e^{ik_n x}}{\sqrt{k_n}} \exp\left(i\tfrac{2n\pi}{D}y\right) + \sum_{m=-\infty}^{\infty} r_{mn} \tfrac{e^{-ik_m x}}{\sqrt{k_m}} \exp\left(i\tfrac{2m\pi}{D}y\right) \qquad (x < R) \\ \Psi_n^l &= \sum_{m=-\infty}^{\infty} t_{mn} \tfrac{e^{ik_m x}}{\sqrt{k_m}} \exp\left(i\tfrac{2m\pi}{D}y\right) \qquad (x > R) \end{aligned} \tag{17}$$

In [10]-(IV.5) the angular momentum decomposition

$$\Psi_n^l(x, y) = \sum_l a_{nl} R_l(r) e^{il\phi} \tag{18}$$

is introduced which is valid for $r < D/2$. The radial wave function is to vanish at the edge of the discs and is therefore determined by the $T$ - matrix of the discs which is diagonal in angular momentum representation:

$$R_l(r) = J_l(kr) \; + \; T_l \, H_l^+(kr). \tag{19}$$

$J_l$ and $H_l^+$ are Bessel and Hankel functions, respectively [11]. Writing the Lippmann - Schwinger equation for the problem in cylindrical coordinates it is possible to derive a system of linear equations determining all the coefficients $a_{nl}$. We will not repeat this here but refer to [10]-(IV.9). However, in our case, where we want to include evanescent modes n is no more restricted by energy conservation but runs over all integer values.

The only new ingredient required by this is a general angular momentum decomposition formula for exponentials:

$$\exp(ik_x x + ik_y y) = \sum_{l=-\infty}^{+\infty} e^{il\phi_{\vec{r}}} r_l J_l(kr) \tag{20}$$

$$r_l = \left( \frac{Sign(l)\, ik_x + k_y}{k} \right)^{|l|} \tag{21}$$

which is valid for both real and imaginary $k_x$ and $k_y$ as long as $k^2 = k_x^2 + k_y^2$ (see appendix A).

For the coefficients in the decomposition of $\exp(-ik_n x + i\frac{n\pi}{L}y)$ we will use the symbol $R_{nl}$ and from (21) we have

$$R_{-nl} = (-1)^l R_{n-l}. \tag{22}$$

Proceeding in the same way which led to [10]-(IV.9) we now obtain

$$\sum_{l'} a_{nl'} \left\{ \delta_{ll'} - F_{l-l'}(kD) T_{l'} \right\} = R_{n-l} \tag{23}$$

with

$$a_{-n\,l} = (-1)^l a_{n\,-l} \tag{24}$$

apparent from symmetry. The structure function $F_L(x)$ comes from the angular momentum decomposition of the Lattice Green function

$$g_k^{(+)}(\vec{r};\vec{r'}) = \sum_{n=-\infty}^{\infty} G_k^{(+)}(\vec{r};\vec{r'} + nD\vec{e_y}) \quad (0 \le |\vec{r'}| \le |\vec{r}| \le D/2) \tag{25}$$

where

$$G_k^{(+)}(\vec{r},\vec{r'}) = i\pi H_0^{(+)}(k|\vec{r} - \vec{r'}|) \tag{26}$$

is the Green function of the plane. The structure function is defined through

$$F_l(x) = 2\sum_{n=1}^{\infty} \cos(\frac{l\pi}{2}) H_l^+(nx) \tag{27}$$

which is zero for odd l. The evaluation of this structure function is the most difficult part of the applied method. It is adapted from [10] where it is described in appendix B.

We take now antisymmetric combinations of the $\Psi_n$ in order to match our boundary conditions and assume in the following $x > 0$:

$$\Psi_n^-(x,y) = \frac{1}{2i}(\Psi_n^l(x,y) - \Psi_n^l(x,-y)) \tag{28}$$

$$\Psi_n(x,y) = \Psi_n^-(-x,y) - \Psi_n^-(x,y) \tag{29}$$

The normalization was chosen to yield (9) for the normal mode decomposition after using the symmetry of the problem to restrict the sum over modes to positive integers $n$, $m = 1, 2, ....$ Now the $S$ - matrix we are interested in is connected to the reflection and transmission amplitudes of the infinite array by

$$S_{mn}^L = (t_{mn} - t_{-mn}) - (r_{mn} - r_{-mn}) \tag{30}$$

The angular representation turns into

$$\Psi_n(x,y) = -\frac{2}{\sqrt{k_n}} \sum_l (a_{nl} - a_{n-l}) R_l(r) \sin l\phi \tag{31}$$

where $l = 2, 4, ...$ is now an index running over positive even angular momenta. From the linear system (23) we derive using the symmetries with respect to inverted indices

$$\sum_l (a_{nl'} - a_{n-l'})(\delta_{ll'} + (F_{l+l'} - F_{l-l'})T_{l'}) \ = \ R_{n-l} - R_{nl}. \tag{32}$$

The angular momentum representation of the wave function in the vicinity of the discs can now be obtained by solving this linear set of equations. However, in order to get the S - matrix we need the normal mode representation of the wave function. For this purpose we use the fact that any two solutions of (2) obey

$$0 \ = \ \int_{\delta\mathcal{G}} d\vec{s}(\Psi_1 \frac{\partial}{\partial \vec{r}} \Psi_2 - \Psi_2 \frac{\partial}{\partial \vec{r}} \Psi_1) \tag{33}$$

for an arbitrary area $\mathcal{G}$. We choose $\mathcal{G}$ to be enclosed by the channel wall $y = L$, the line $x = L$ and the circle around the origin with radius $L$ (cmp. Fig. 3). For the wave functions we take $\Psi_1 = \Psi_n$ and $\Psi_2 = \sin k_{m'}x \ \sin \frac{m'\pi}{L} y$. They both are antisymmetric with respect to $y = L$, therefore this line does not contribute to the integral and we obtain $I_\phi + I_y = 0$ with

$$\begin{aligned} I_y \ &= \ \int_0^L dy \left( \Psi_1 \frac{\partial}{\partial x} \Psi_2 - \Psi_2 \frac{\partial}{\partial x} \Psi_1 \right)_{x=L} \\ &= \ \sqrt{k_{m'}} (\delta_{m'n} - S_{m'n}^L) \end{aligned} \tag{34}$$

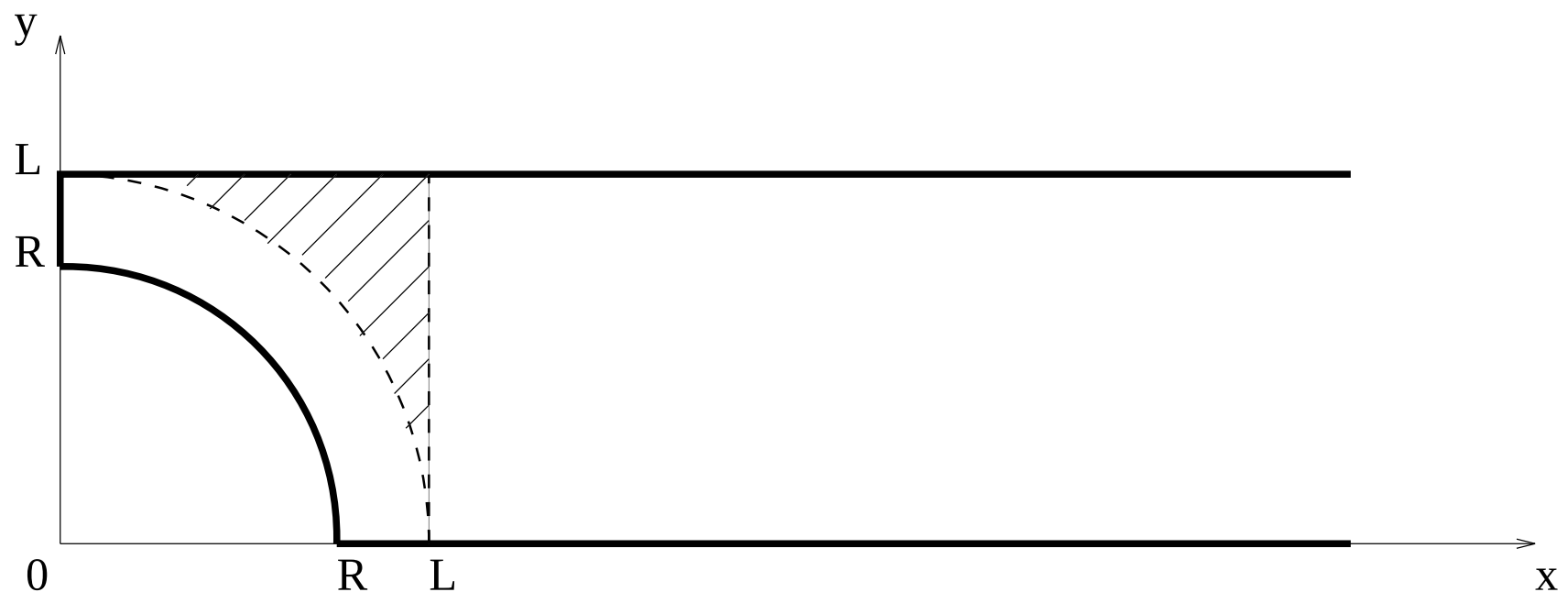

Figure 3: *The desymmetrized section, the area $\mathcal{G}$ is hatched.*

$$
\begin{aligned}
I_\phi &= \int_0^{\pi/2} L d\phi \left( \Psi_1 \frac{\partial}{\partial r} \Psi_2 - \Psi_2 \frac{\partial}{\partial r} \Psi_1 \right)_{r=L} \\
&= \sum_l \frac{T_l}{\sqrt{k_n}} (a_{nl} - a_{n-l})(R_{m'l} - R_{m'-l}) \qquad (35)
\end{aligned}
$$

From these two formulas we extract $S^L$:

$$S^L_{mn} = \delta_{mn} + \frac{1}{\sqrt{k_n k_m}} \sum_l T_l (a_{nl} - a_{n-l})(R_{m'l} - R_{m'-l}) \qquad (36)$$

For a compact notation we introduce the matrices:

$$A_{nl} = \frac{2i}{\sqrt{k_n D}} (a_{nl} - a_{n-l}) T_l \qquad (37)$$

$$C_{nl} = \frac{2i}{\sqrt{k_n D}} (R_{nl} - R_{n\,-l}) \qquad (38)$$

$$P_{ll'} = \frac{\delta_{ll'}}{T_{l'}} + F_{l+l'} - F_{l-l'} \qquad (39)$$

(37) and (36) take now the forms

$$A\,P = -C \qquad (40)$$

$$S^L = I - C\,A^T \qquad (41)$$

Putting them together we end up with our final expression for $S^L$:

$$S^L = I + C\,P^{-1}\,C^T. \qquad (42)$$

The above relations give explicit expressions for $S^L$ in terms of known quantities. They form the basis for the numerical evaluation of $S^L$.

# 3 Numerical Results

We start the presentation of our numerical results with an investigation of the accuracy of the eigenvalues we calculated. For this purpose we employ the symmetry of the billiard with respect to reflections from the diagonal $x = y$. In the scattering approach this symmetry is broken by attaching the channel to one of the the billiard walls. However, the eigenvalues $k_i$ of the billiard must belong to one of the two representations of the symmetry. So they can be divided into two subgroups $k_i^+$ and $k_i^-$ with the wave function inside the billiard symmetric or antisymmetric, respectively:

$$\Psi^{k_i^\pm}(x, y) = \pm\Psi^{k_i^\pm}(y, x) \tag{43}$$

In polar coordinates the symmetric wave functions will contain only angular momenta $l = 4i$, $i = 1, 2, \ldots$ If we denote the projection onto this subspace by $Q^+$ and the projection onto the complementary set of angular momenta by $Q^-$ we have therefore in terms of the matrices (37)

$$A(k_i^+) = Q^+ A(k_i^+) \qquad 0 = Q^- A(k_i^+) \tag{44}$$

Using $S$ - matrices calculated with restricted space of angular momenta

$$S^{L\pm} = I - CQ^\pm A^T \tag{45}$$

we have thus at the symmetric eigenvalues $k_i^+$ besides

$$Z(k_i^+) := \det(I - S^R S^L(k_i^+)) = 0 \tag{46}$$

also

$$Z^+(k_i^+) \;:=\; \det\Big(I - S^R S^{L+}(k_i^+)\Big) \;=\; 0 \tag{47}$$

Though we cannot make use of the symmetry for a simplification of the calculation of eigenvalues (e. g. by searching only for all the $k_i^-$ as in [1] and [3]), we obtain at least a means for estimating the deviations of the eigenvalues $\overline{k_i^+}$ computed by searching the zeros of (46) from the correct values $k_i^+$: The numerical errors will cause deviations of $Z^+(\overline{k_i^+})$ from zero and the corresponding shift in the eigenvalue reads

$$\delta k_i = \left|\overline{k_i^+} - k_i^+\right| \sim \left|\frac{Z^+}{\frac{d}{dk} Z^+|_{\overline{k_i^+}}}\right| \tag{48}$$

Obviously, analogous reasoning holds for the $k_i^-$.

It is appropriate to measure the error estimate in units of the mean level spacing around the considered eigenvalue. The circles in fig. (4) show this relative error in a logarithmic plot for two different radii of the circle[1]. The error is about $10^{-6}$ for a quite small circle with $R = 0.4$ and about $10^{-5}$ for a very large circle ($R = 0.9$). We conclude that the scattering approach is well suited for an accurate calculation of thousands of eigenvalues for any given radius. The crosses in fig. (4) demonstrate the relative error in the generalized symmetry and unitarity relations for $S^L$ at the eigenvalues. They were obtained by picking the maximum relative deviation of a single matrix element of $S^L$ from the prediction of (80) - (82) and (89). We see that the violation of these relations is about $10^{-12}$ and thus much smaller than the deviations in the eigenvalues. This means, that the most important error is introduced through the imaginary part of the structure function (27) which does not influence the checked relations (compare appendix B).

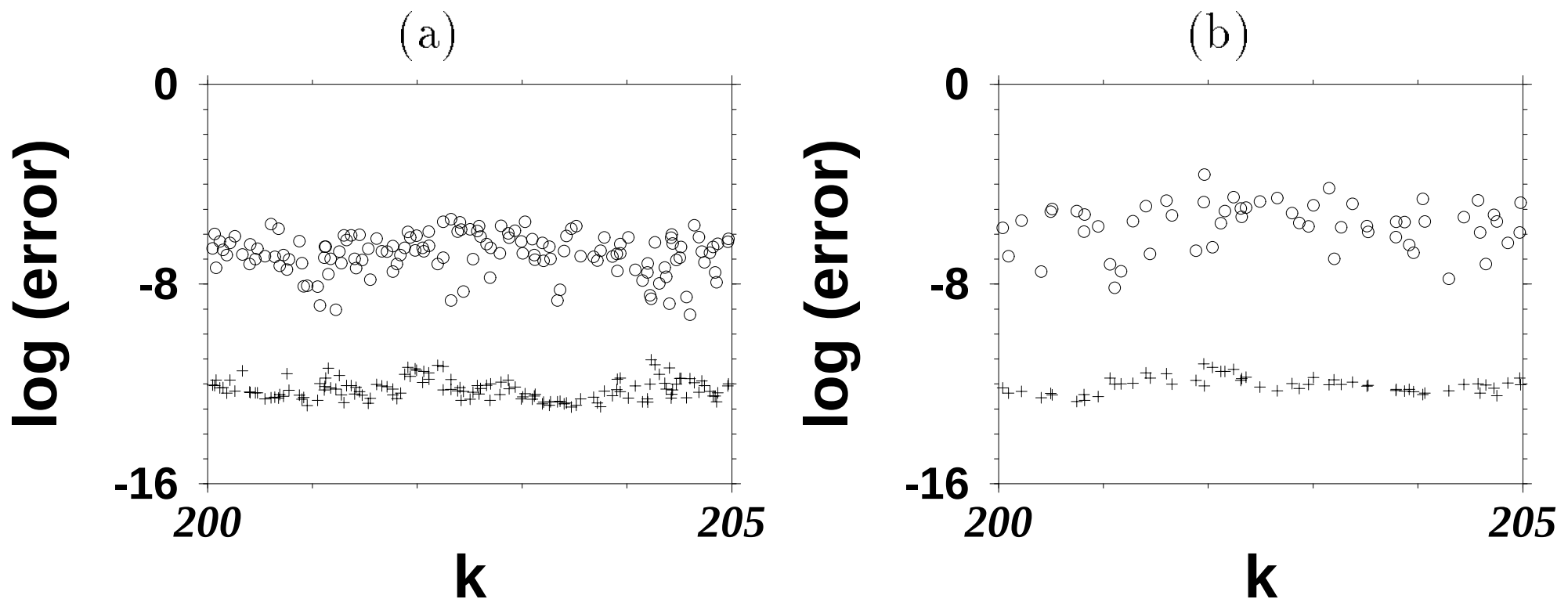

Figure 4: *Estimate of the numerical error at some eigenvalues for (a)* $R = 0.4$ *and (b)* $R = 0.9$ *The eigenvalues were obtained including 4 closed channels. The circles show the estimated deviation in the wave number* $\delta k_i$ *normalized to the mean level spacing. The crosses denote the corresponding violation of the symmetry relations for the generalized S.*

The relative deviations of the eigenvalues are expected to increase with growing radius of the circle and growing energy, since both will cause more angular momenta to contribute to the decomposition of the wave function around the circle:

$$l_{max} \sim Rk \tag{49}$$

and the structure function in turn is needed up to $2l_{max}$. Thus the maximum wave number $k_{max}$ where the relative deviation of eigenvalues is well below the mean level spacing decreases with growing radius of the disc. The number of eigenvalues up to $k_{max}$ decreases even faster, since growing radius of the inside circle means shrinking area of the billiard. For example, at $R = 0.4$ it is possible to obtain about 5,000 eigenvalues for each of the two symmetry classes whereas for $R = 0.75$ we are restricted to the lowest 3,000.

[1] In this and in all the following plots the edge length of the square was fixed at $L = 1$.

Once we have established the accuracy of the method, we use the resulting data to check various statistical properties of the spectrum.

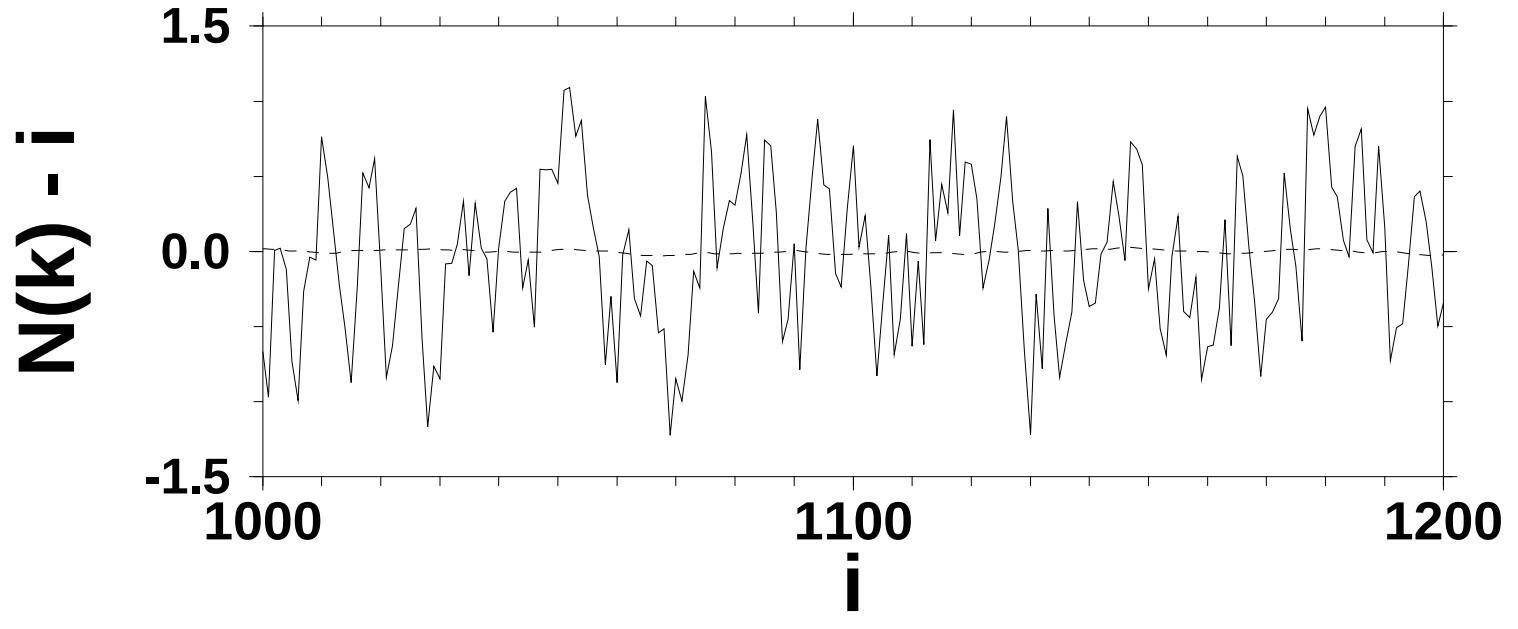

Figure 5: *Deviation of the eigenvalues from Weyl's law.* $\overline{N}(k_i) - i$ *at 200 eigenvalues for* $R = 0.75$*. The dashed line is a running average over 200 eigenvalues.*

The spectrum of eigenvalues $k_i$ of the billiard can be described by the spectral density

$$d(k) = \sum_{i=1}^{\infty} \delta(k - k_i). \tag{50}$$

The integral of the spectral density

$$N(k) = \int_0^k dk' \; d(k') \tag{51}$$

increases by one at each eigenvalue and is called spectral staircase. The mean density of the eigenvalue spectrum in an interval containing many levels obeys a generalization of Weyl's law [14] which reads in terms of the spectral staircase

$$\overline{N(k)} = \frac{A}{4\pi} k^2 - \frac{u}{4\pi} k + C. \tag{52}$$

$A$ is the area of the billiard, $u$ the length of the circumference and C a constant which can be obtained from geometry as well. The oscillations of the actual spectral staircase $N(k)$ around this mean are shown in fig. (5) for a fraction of the spectrum containing 200 eigenvalues. The mean value of 200 neighboring eigenvalues is plotted with a dashed line and is almost constant zero.

The oscillations of $N(k) - \overline{N(k)}$ can be described semiclassically using the periodic orbits of the system. Besides unstable and isolated orbits the SB also permits orbits which are neutral and form continuous families. These so called bouncing ball orbits are not reflected from the quarter circle but only from the straight walls of the billiard. In [15] it is shown how the contributions of both unstable and bouncing ball orbits can be used to obtain a semiclassical prediction for the length spectrum which is a finite Fourier transform of the oscillating part of the spectral density:

$$D(x, k_{max}) = \frac{1}{k_{max}} \int_0^{k_{max}} dk \;\cos kx \left[d(k) - \overline{d(k)}\right] \tag{53}$$

This function is plotted in fig. (6) using 3350 eigenvalues for a SB with $R = 0.75$ The lengths of the shortest periodic orbits are shown in fig. (6) by vertical lines. Each of these lengths corresponds to a peak in the spectrum. For the given radius, the only possible family of bouncing ball orbits is formed by orbits reflected back and forth between opposite sides of the billiard. These orbits have the length $x = 2$ and cause the most prominent peak in the length spectrum. A semiclassical prediction for the length spectrum was obtained using input from just the shortest ten periodic orbits and is displayed with a dashed line. It is a good approximation to the Fourier transform of the eigenvalue density.

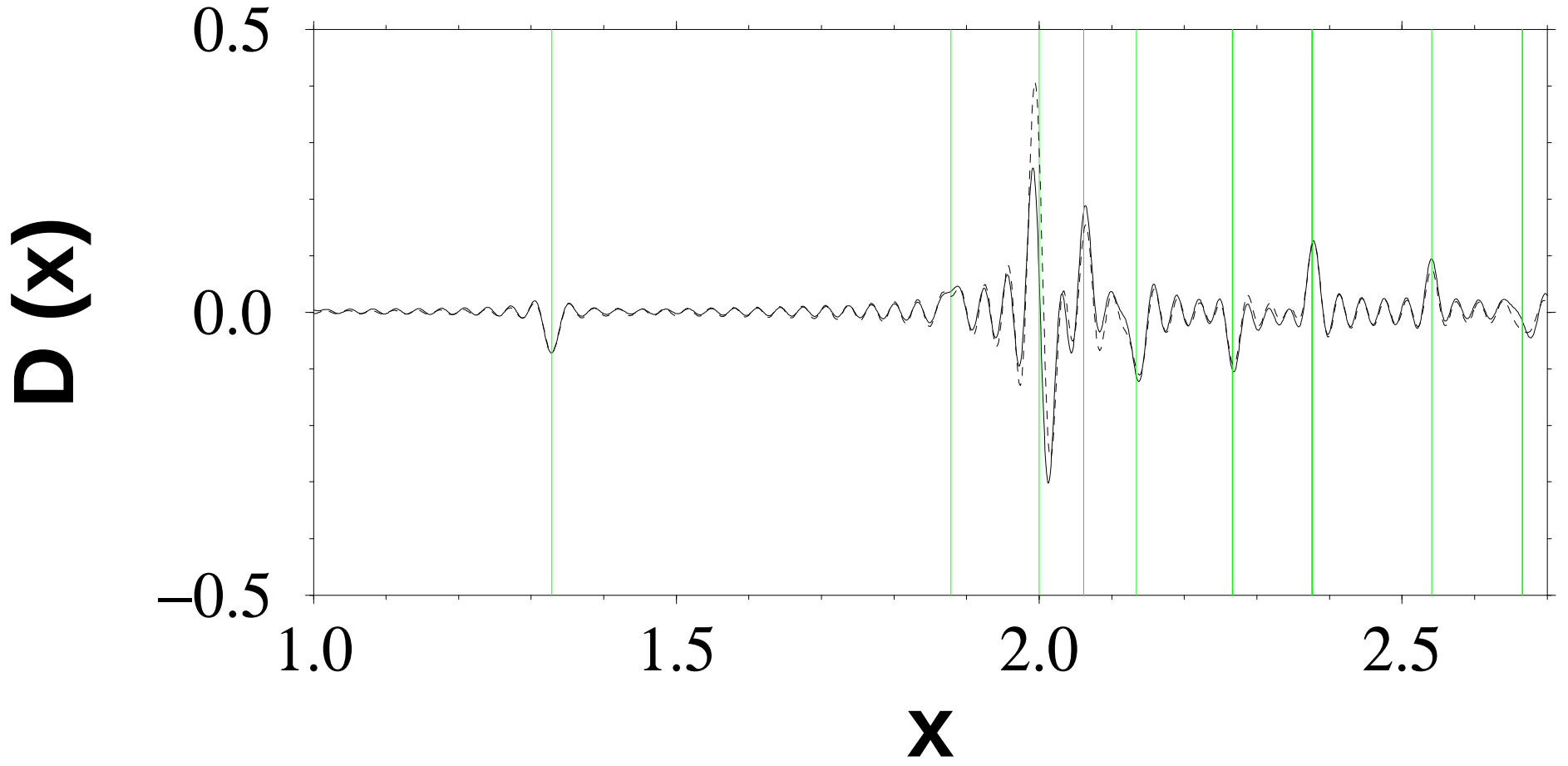

Figure 6: *Length spectrum obtained from 3350 eigenvalues for* $R = 0.75$ *and* $k_{max} = 200$ *(full line) versus a semiclassical prediction containing the shortest 10 periodic orbits of the billiard (dashed line). The vertical lines denote the lengths of the used periodic orbits.*

The SB is known to exhibit classically chaotic dynamics for any positive value of the inside radius $R$. Therefore the eigenvalue spectrum of the billiard can be described by the Gaussian orthogonal ensemble of random matrices GOE. To check this we employ the statistics of spacings between neighboring eigenvalues and the number variance $\Sigma(L)$ and rigidity $\Delta_3(L)$ which contain the two point correlations in the spectrum. The spectrum is unfolded using the generalized Weyl's law and the statistics are compared to the predictions of random matrix theory([16], [17]).

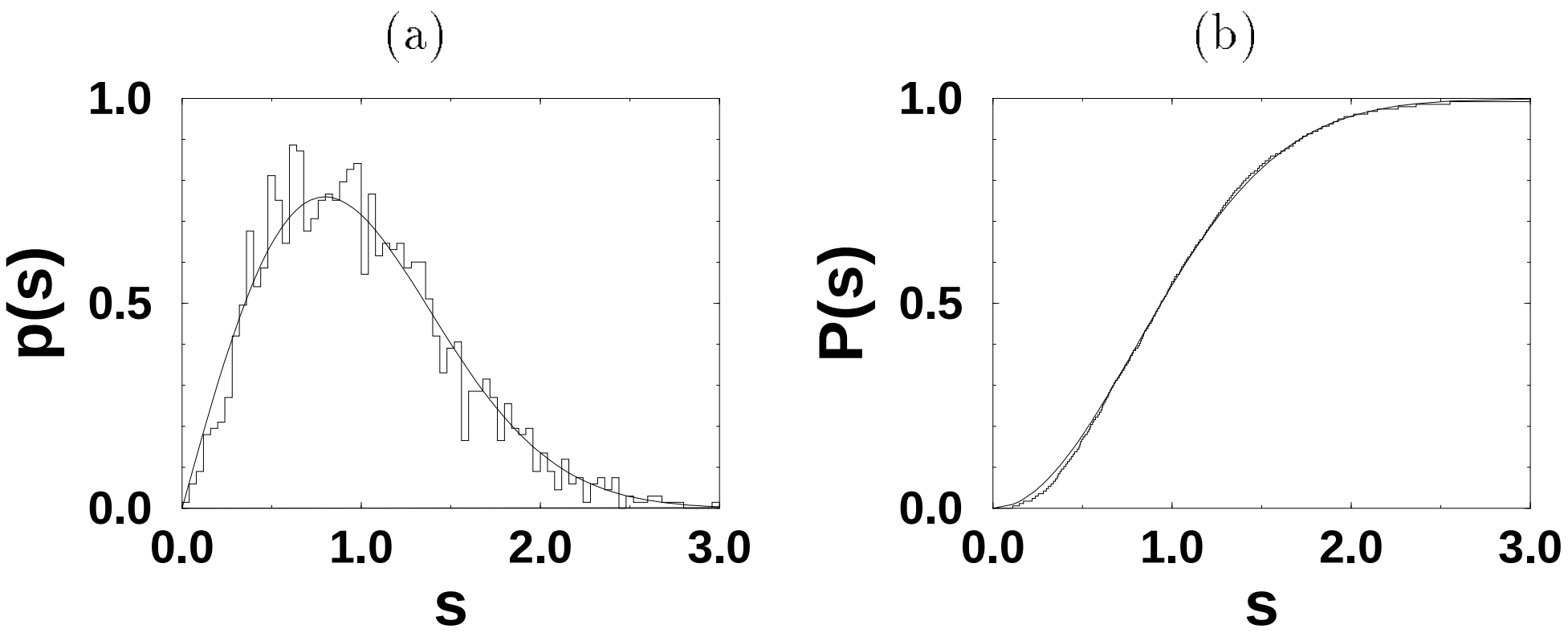

Figure 7: *1600 antisymmetric eigenvalues for* $R = 0.75$ *(a) Nearest neighbor spacing distribution (b) integrated distribution. The smooth line is the prediction from GOE.*

Fig. (7a, b) shows that the nearest neighbor spacings of the eigenvalue spectrum are in good agreement with the GOE. The two point measures, however, follow the random matrix prediction only up to a maximum length and saturate above it. This behavior has been explained by Berry [18] using semiclassical arguments (see fig. (8)).

In [10] and [12] it was shown that a S - matrix corresponding to a classically chaotic scattering system with time reversal symmetry can well be described by the circular orthogonal ensemble (COE) of random matrices. We would like to check some of the predictions of random matrix theory with the unitary part of the S - matrix employed for the billiard quantization.

Consider first the distribution of the eigenphases of $S^L$ on the unit circle. For COE we expect a constant density $\frac{1}{2\pi}$. On the other hand $S^L$ approaches $I$ as the the radius shrinks to zero and for any radius we have a contribution to S from direct reactions. Therefore we have a peak in the eigenphase density around zero which falls in magnitude for increasing radius (fig. 9). We are not interested, however, in the contributions from direct reactions and will use therefore $S^R S^L$ where they are not present. Fig. (10) shows that the corresponding eigenphase distribution is constant $\frac{1}{2\pi}$.

The distribution of spacings between neighboring eigenphases is expected to undergo a transition from an exponential behavior $p(s) = e^{-s}$ to the Wigner surmise $p(s) = \frac{\pi}{2} s e^{-\frac{\pi}{4}s^2}$ as the dynamics of the corresponding classical system changes from regular to chaotic. Indeed, fig. (11) shows a distribution close to the Wigner surmise for large radius and a Poisson - like behavior for small discs.

Two point correlations in the spectrum of eigenphases provide another test of the random matrix hypothesis for for the $S$ - matrix. They can be expressed in terms of the probability density $P_2(\eta)$ of finding two eigenphases with a distance $\eta$. The Fourier coefficients $s_N^{(\Lambda)}$ of this density are related to the eigenvalues [12, 13] through

$$\begin{aligned} s_N^{(\Lambda)} &:= (\Lambda - 1) \int_0^{2\pi} d\eta \, P_2(\eta) \, e^{i\eta N} \\ &= \frac{1}{\Lambda} |Tr \; S^N|^2 - 1 \end{aligned} \tag{54}$$

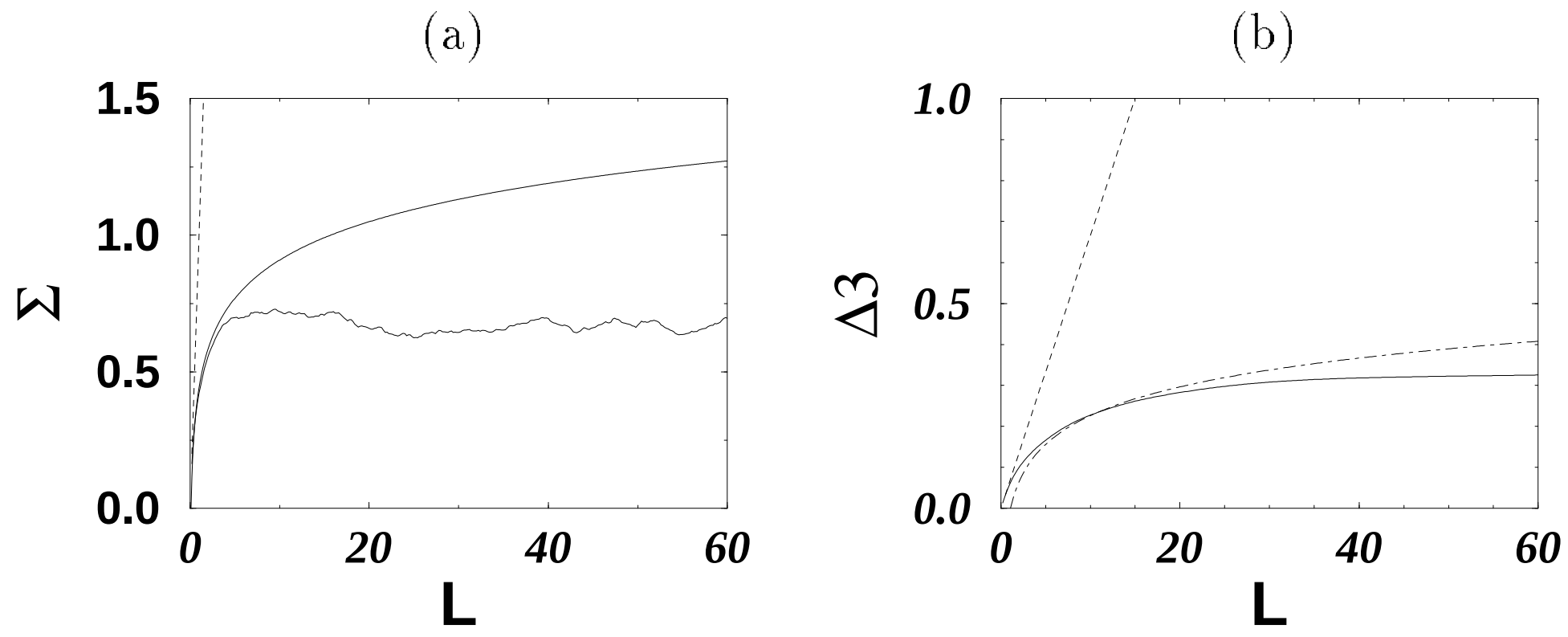

Figure 8: *1600 antisymmetric eigenvalues for* $R = 0.75$ *(a) Number variance (b)* $\Delta_3$ *statistics. The dashed - dotted line denotes the GOE prediction, the dashed line corresponds to a Poisson process.*

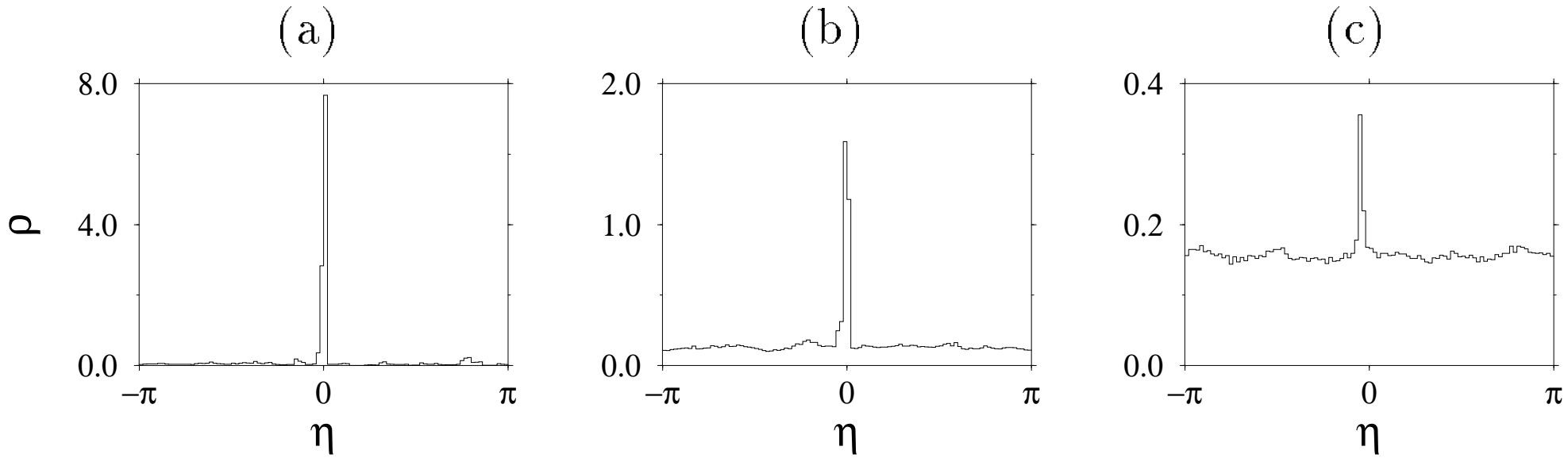

Figure 9: *Density distribution of the eigenphases of* $S^L$ *for (a)* $R = 0.4$*, (b)* $R = 0.65$*, (c)* $R = 0.9$ *Energy average over 400 matrices with 50 open channels.*

A semiclassical description of the behavior of the coefficients $s_N^{(\Lambda)}$ in agreement with the COE prediction [16] has been derived in [12]. Apart from a non generic region for very small N one expects $s_N^{(\Lambda)} \sim \frac{2N}{L} - 1$ for $N^* < N < \Lambda$ and $s_N^{(\Lambda)} \sim 0$ for $N \gg \Lambda$. From fig. (12a - c) one can see that this prediction describes the $S$

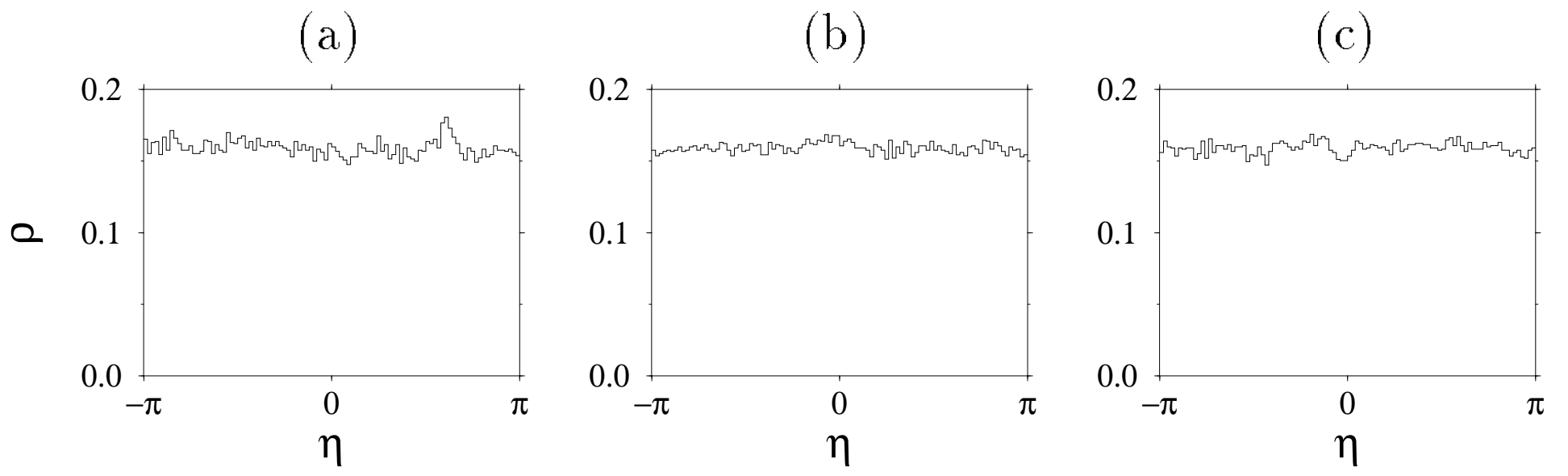

Figure 10: *Density distribution of the eigenphases of* $S^R S^L$ *for (a)* $R = 0.4$*, (b)* $R = 0.65$*, (c)* $R = 0.9$ *Energy average over 400 matrices with 50 open channels.*

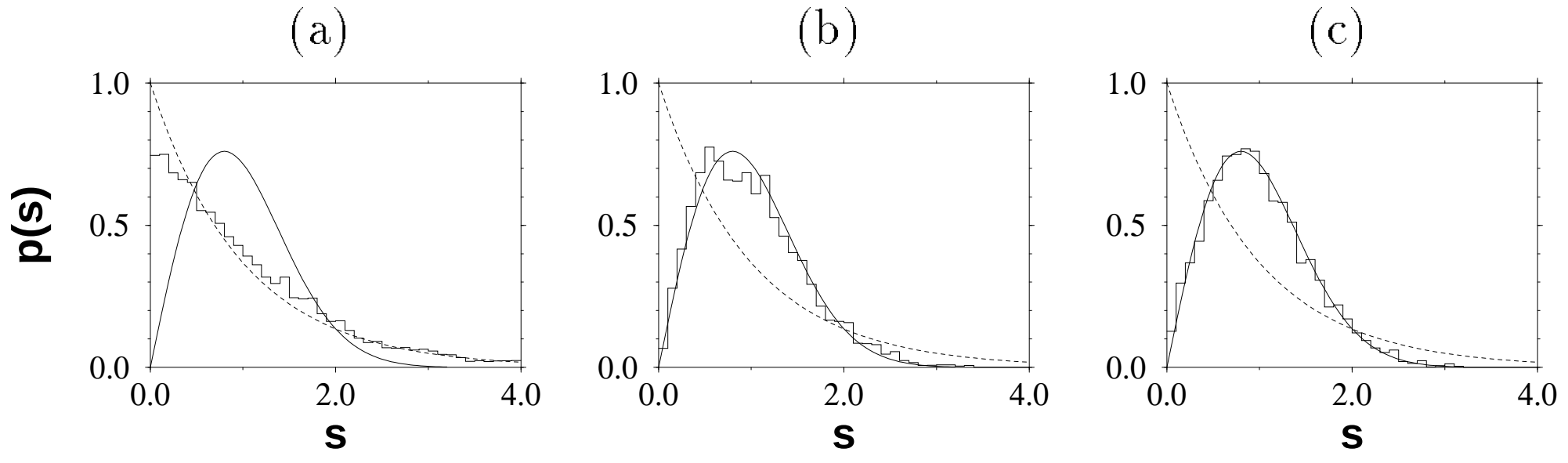

Figure 11: *Nearest neighbor spacing distribution of the eigenphases of* $S^R S^L$ *for (a)* $R = 0.4$*, (b)* $R = 0.65$*, (c)* $R = 0.9$ *Energy average over 400 matrices with 50 open channels.*

- matrix of the SB with large radius very well. For small radius the S - matrix can not be described by random matrix theory.

# 4 Semiclassical Quantization

As was explained in the introduction, the semiclassical approximation is derived in two steps. In the first, we restrict the $S$ matrix in the secular equation (10) to the space of open channels. This is the "semiquantal" approximation for the secular equation, since the restricted $S$ matrix is the fully quantal $S$ matrix which describes the reflection as measured at large distances. This restriction is a natural step within the semiclassical theory, because evanescent modes correspond to propagation with imaginary momenta, and generically their contribution is exponentially small in $\hbar$. However, in the present case, the imaginary momenta start at zero at threshold, so that for any finite value of $\hbar$, there exists a corresponding neighborhood of the threshold energies, where the decay of the evanescent mode which is about to be opened is too slow to justify its elimination. To check the accuracy of the semiquantal approximation, we used the methods and numerical data which were discussed in Chapter 3. above.

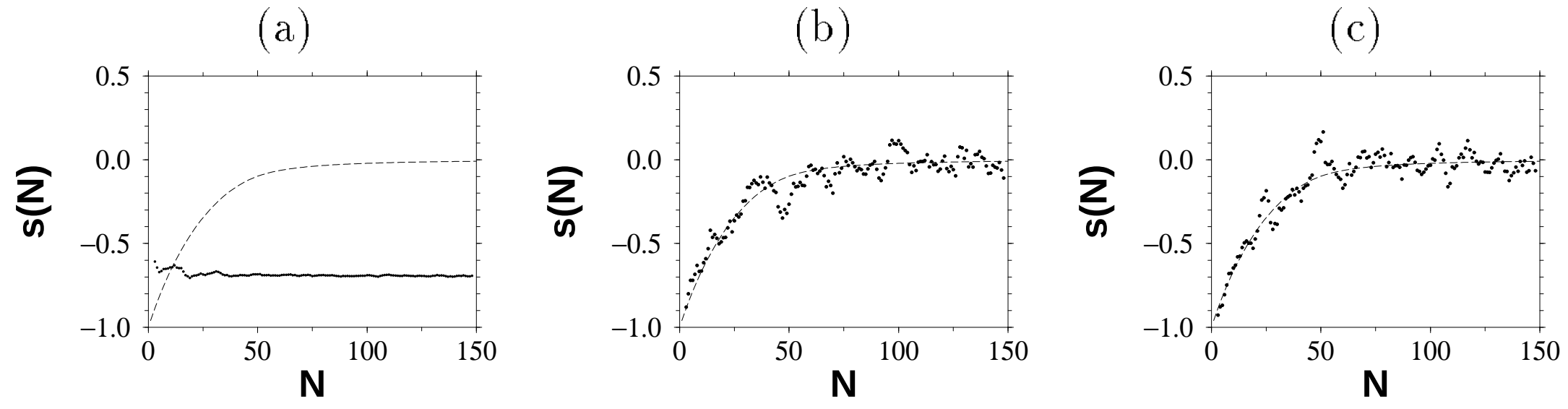

Figure 12: *Fourier coefficients* $s_N^{(\Lambda)}$ *of the cluster function as a function of N. (a)* $R = 0.4$, *(b)* $R = 0.65$, *(c)* $R = 0.9$ *Energy average over 400 matrices with 50 open channels. The dashed line shows the prediction of the COE.*

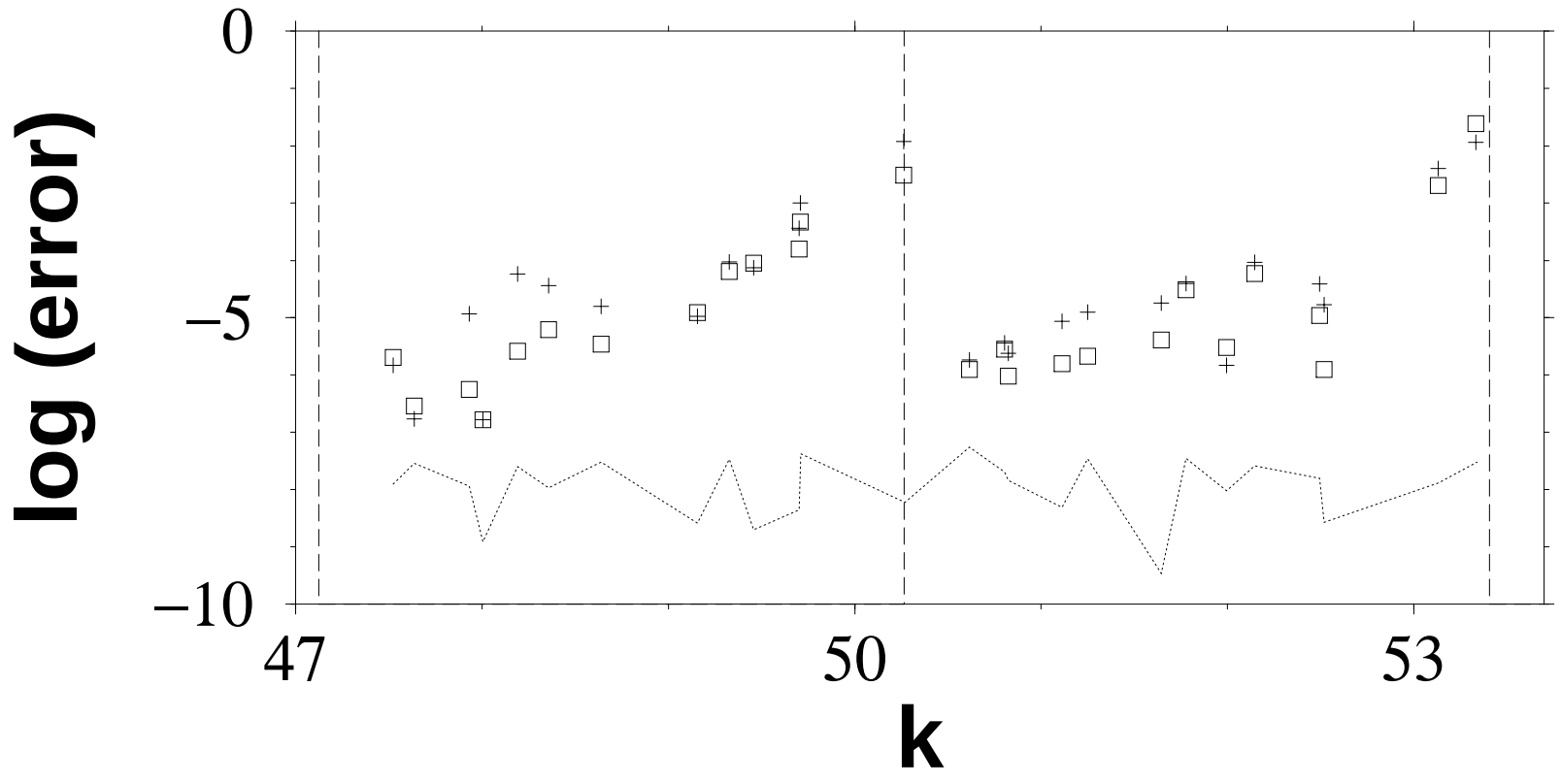

Figure 13: *Estimation of the numerical error* $\delta k_i$ *for* $R = 0.75$ *The eigenvalues were obtained including 2 closed modes (full line) and no closed modes (squares). The crosses show the difference between both sets of eigenvalues. All data are in units of the mean level spacing.*

In fig. (13) we plotted all eigenvalues between the wave numbers where the 15th and the 17th channel open. These thresholds are marked with vertical dashed lines. The full line shows the estimated deviation between the eigenvalues calculated with two closed channels and the exact locations of eigenvalues. It has been checked that inclusion of more evanescent modes does not alter the numerical eigenvalues for the parameters under consideration but neglecting these two lowest closed modes does: The squares show the deviations of the eigenvalues obtained with unitary $S$ from those with 2 closed channels and the crosses mark the corresponding error estimate according to (48). Both are of the same order of magnitude as it should be if (48) is correct. We see that the error coming from neglecting evanescent modes is in general not very large. It rises exponentially with decreasing distance to the next threshold where a new channel starts to conduct as predicted in [5].

The same behavior is also apparent from fig. (14). Here, the estimated relative error is shown for two different radii. The dashed line corresponds to the values obtained from 2 closed channels, squares and crosses mark the results for 0 and 1 closed channels, respectively. For the lower radius $R = 0.75$ there is no difference between the values for 1 and 2 closed channels but for large radius higher evanescent modes can also contribute as one can see from fig. (14b). For large radius the influence of the evanescent modes is in general bigger. This can be understood since the evanescent modes decay exponentially in the channel region $R < x < L$ and should contribute less when the distance for the decay increases. In [5] the situation is different. There the channel was attached at the opening between the circle and the billiard wall. If the radius grows the distance for the decay increases and the influence of evanescent modes must diminish.

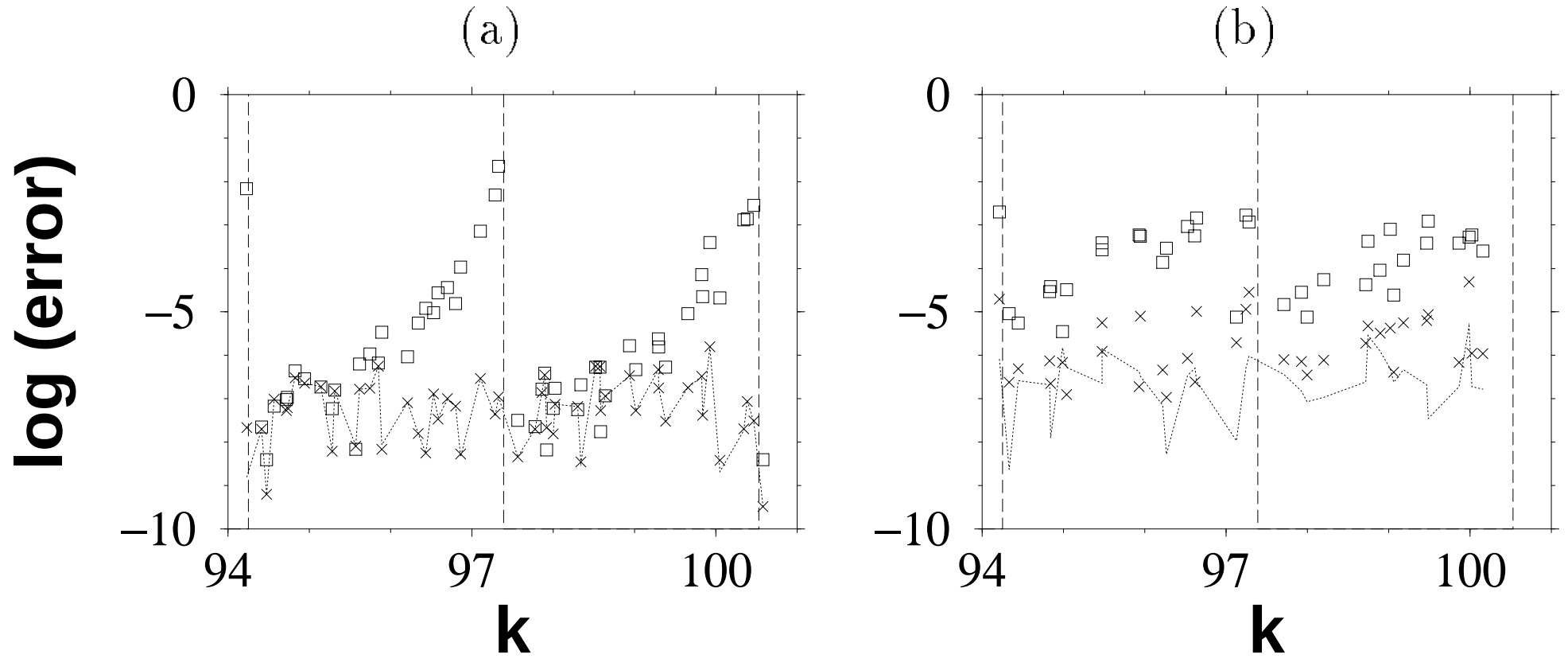

Figure 14: *Estimation of the numerical error $\delta k_i$ for (a) $R = 0.75$, (b) $R = 0.9$ The eigenvalues were computed with 0 (squares), 1 (crosses) and 2 (line) evanescent modes and the error is in units of the mean level spacing.*

In [5] it is proposed that the error from neglecting closed channels should scale with the normalized absolute value of the wave number of the first evanescent mode

$$\delta = \frac{|k_{\Lambda+1}|L}{\pi\sqrt{2\Lambda+1}} \tag{55}$$

which is a number between 0 and 1. In fig. (15) we show the deviations of 1000 eigenvalues computed without closed channels from those obtained with one closed channel versus $\delta$. The error increases exponentially as the magnitude of the lowest closed mode falls to zero. The error will be of the order of the mean level spacing if one eigenvalue is very close below a channel opening. The eigenvalues are, however, concentrated in regions where the error is very small.

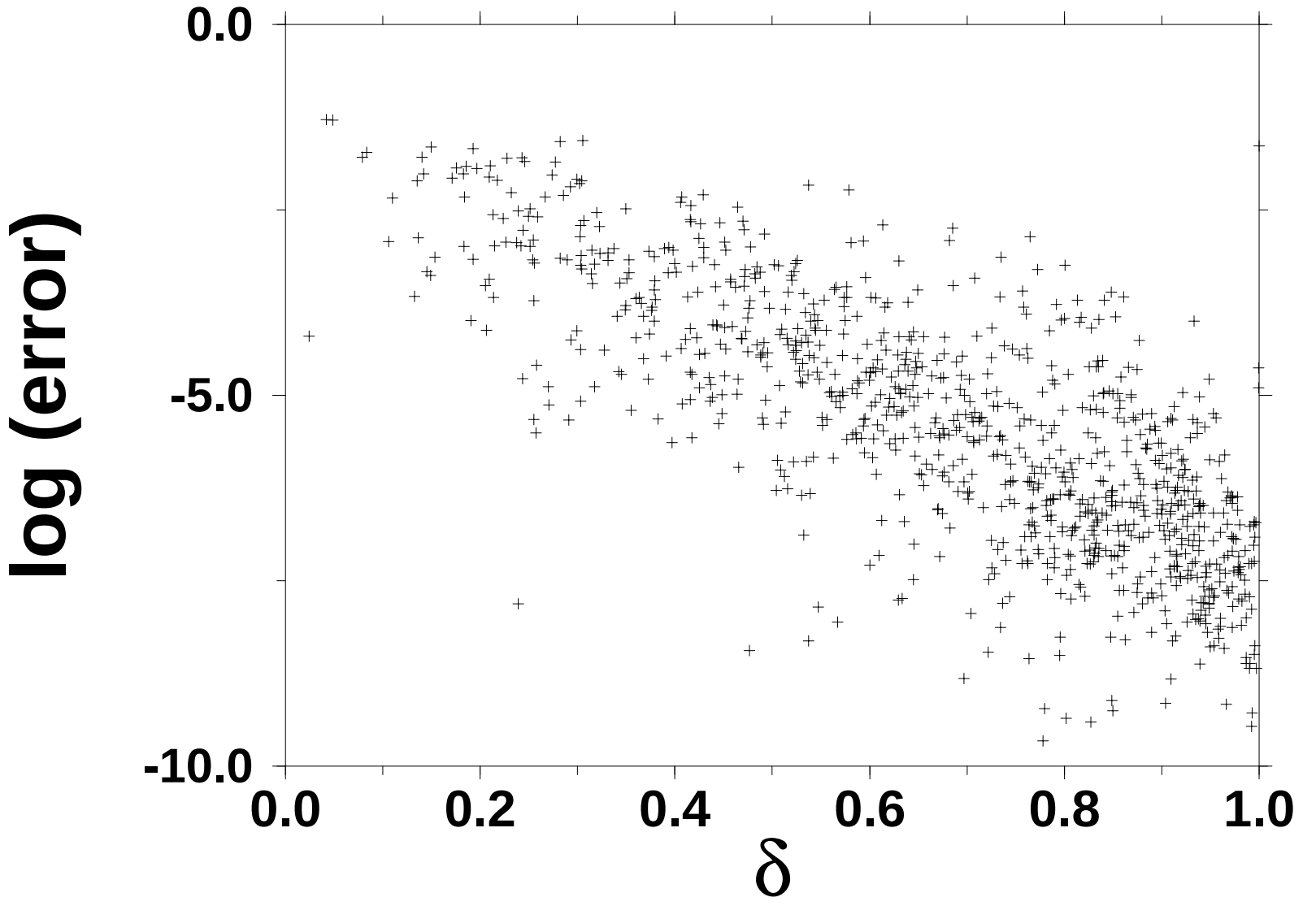

Figure 15: *Deviations between eigenvalues obtained with no and with one closed channel. First 1000 eigenvalues of the billiard with* $R = 0.75$ *The errors are normalized to the mean level spacing and the abscissa is the magnitude of* $k_{\Lambda+1}$ *in units of* $\pi/d(2\Lambda+1)^{1/2}$.

To end the discussion of the influence of evanescent modes we show the behavior of the eigenvalues $S_i$ of $S^R S^L$ in the vicinity of a channel opening. In fig. (16a) the magnitude of all the eigenvalues of $S^R S^L$ is shown around the threshold for the opening of the second channel which is marked with a vertical line. Below the threshold there is only one open channel and therefore just one eigenvalue of $S^R S^L$ is on the unit circle. Far away from the threshold the magnitude of all the other eigenvalues is very small and therefore they can be neglected. When $k$ approaches the threshold the magnitude of the lowest evanescent mode increases from 0 to 1. For the same energy region in fig. (16b) the behavior of the eigenvalues is shown in the complex plane. All the eigenvalues corresponding to closed channels are located in the origin. There is one eigenvalue corresponding to the first open channel traveling at constant speed on the unit circle. The eigenvalue corresponding to the second open channel tends to 1 as the energy approaches the threshold. This means, that the secular equation (1) will have additional solutions at the channel openings which do not correspond to billiard eigenenergies. The derivative of the eigenvalue of the opening channel is discontinuous at the threshold.

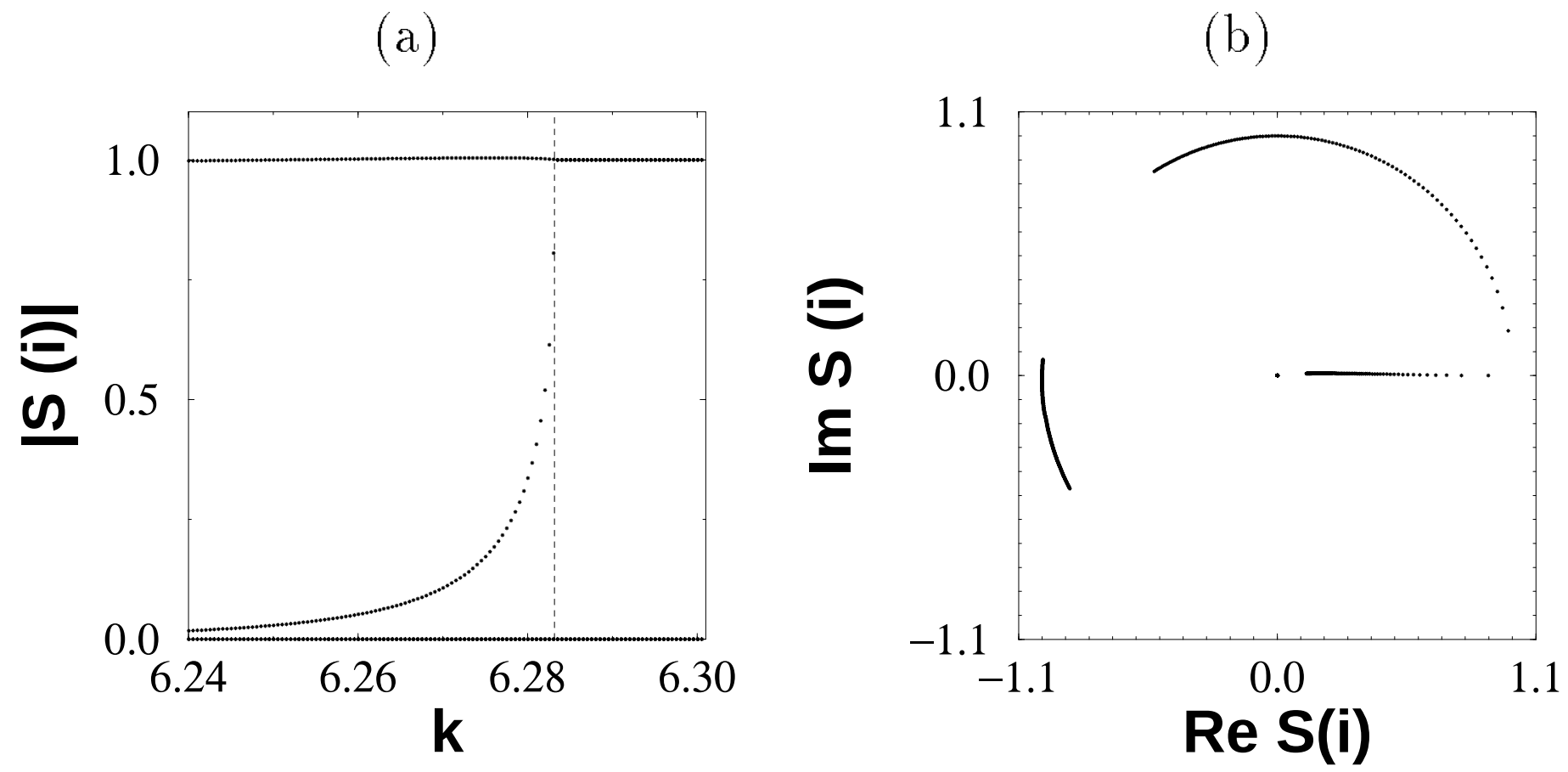

Figure 16: *Behavior of the eigenvalues of $S^R S^L$ for $R = 0.75$ in the vicinity of the opening of the second channel (vertical line). (a) magnitude (b) location in the complex plane*

Now that we have assessed the quality of the semiquantal approximation, we shall proceed to discuss some of its properties. The semiquantal secular equation reads

$$Z_{sq}(k) = \det(I - S(k)) = 0, \tag{56}$$

where the dimension of $S(k)$ is $\Lambda(k) = [\frac{kL}{\pi}]$. It can be written as

$$Z_{sq} = \sum_{l=0}^{\Lambda} f_l(k) \tag{57}$$

in terms of $f_l(k)$, the coefficients of the characteristic polynomial of $S(k)$. The unitarity of $S(k)$ implies the important symmetry

$$\exp[-i\Theta(k)/2] f_l = \exp[i\Theta(k)/2] f^*_{\Lambda-l} \tag{58}$$

where $\Theta(k) = \sum_{l=1}^{\Lambda} \theta_l(k) - \Lambda\pi$, and $e^{i\theta_l(k)}$ are the eigenvalues of $S(k)$. The quantity $\Theta(k)/(2\pi)$ can be shown to give the smooth number function of the spectrum. This follows from the observation that for $k$ values away from thresholds, the spectral density is given by

$$d(k) = \sum_{l=1}^{\Lambda} \delta_p(\theta_l(k)) \frac{\partial \theta_l(k)}{\partial k} = \frac{1}{2\pi} \frac{\partial \Theta(k)}{\partial k} + \frac{1}{\pi} \sum_{n=1}^{\infty} \frac{1}{n} \frac{\partial \Im Tr(S^n(k)))}{\partial k} \tag{59}$$

The first term on the rhs is the smooth part of the density. Another important message of this expression is that the fluctuating part of the spectral density depends exclusively on $Tr(S^n(k))$. The important rôle of $TrS^n$ is even better emphasized when Newton's identities are used to express the coefficients $f_l$,

$$TrS^m + f_{\Lambda-1} TrS^{m-1} + \ldots + f_{\Lambda-m+1} TrS + m f_{\Lambda-m} = 0; \quad m = 1, \ldots, \Lambda. \tag{60}$$

We conclude, therefore, that the secular equation and hence the spectrum are determined once $TrS^n(k)$ are given for $1 \le n \le [(\Lambda+1)/2]$. The second step in the derivation of the semiclassical quantization is to express these quantities in terms of classical periodic orbits.

To get the semiclassical expression for $TrS^n(k)$, we shall first calculate $S_L$ in the semiclassical approximation, the starting point being equation (14) of chapter (2). Consider first the free Green function

$$G_0^+(xy, x'y') = \frac{i}{2} \sum_{m=1}^{\infty} \frac{\phi_m(y)\phi_m(y')}{k_m} (e^{ik_m|x-x'|} - e^{ik_m(x+x')}) \tag{61}$$

Applying Poisson summation and performing the integrals by the stationary phase approximation, one gets the semiclassical expression

$$G^+_{0,scl}(\vec{r}, \vec{r}\,') = -\frac{e^{i\frac{\pi}{4}}}{(8\pi)^{1/2}} \sum_s \frac{1}{(kl_s(\vec{r}, \vec{r}\,'))^{1/2}} e^{i(kl_s(\vec{r},\vec{r}\,')+\pi\nu_s)} \tag{62}$$

where $s$ is an index which counts the classical paths from $\vec{r}$ to $\vec{r}\,'$ which undergo on their way $\nu_s$ reflections from the boundaries at $x = 0$ , $y = 0$ and $y = L$. Note that the number of reflections from the line $x = 0$ can be at most one. $l_s$ is the length of the path. It should be noted that there exist also complex solutions to the saddle point conditions. They correspond to propagation with imaginary momenta, which, as explained above, are neglected in the present level of approximation.

The semiclassical expression for $G_0^+$ is now substituted in (15) and each of the terms in the Born series is evaluated by performing the integrals by the stationary phase approximation. A tedious but straight forward algebra yields an expression in terms of classical trajectories. To understand the conditions which these trajectories should satisfy we should recall that the channel quantum numbers $m$ are the quantized classical action variables (measured in units of $\hbar$):

$$I = \frac{|k_y|D}{\pi} \tag{63}$$

where $k_y$ is the transverse momentum. The conjugate angle variables are

$$\phi = \frac{\text{Sign}(k_y)y\pi}{D}. \tag{64}$$

The relevant trajectories emerge from the line at $x = L$ with $I_{initial} = m$ and return to $x = L$ with $I_{final} = m'$. In terms of these trajectories one gets:

$$S^L_{mm'}(k) - \delta_{mm'} \approx \frac{1}{(2\pi i)^{1/2}} \sum_s \frac{1}{(|\frac{\partial I_f}{\partial \phi_i}|_s)^{1/2}} e^{i(k\tilde{l}_s+\pi\nu_s)} \tag{65}$$

Where the sum is over all the trajectories which satisfy the boundary conditions specified above. $\tilde{l}_s$ is the reduced length :

$$\tilde{l}_s = l_s - k_m L - k_{m'} L + \phi_s m + \phi_s' m' \tag{66}$$

with $\phi_s$ and $\phi_s'$ the initial and final angle variables (conjugate to $m$ and $m'$). $l_s$ is the total length of the trajectory $s$, and $\nu_s$ is the number of reflections from the boundaries.

Equation (65) plays a central rôle in the semiclassical discussion. Its right hand side looks formally as the standard semiclassical expression for a unitary operator. (See e.g. [9]). The unitarity can be checked by performing the matrix multiplication by the stationary phase approximation, and therefore the rhs of (65)is "semiclassically unitary". On the lhs of (65) we find $S - I$ which becomes unitary only if the identity operator is added to it. The semiclassical approximation tends to miss this term (compare e.g. the case of elastic scattering from a hard disc in the plane. There the semiclassical $T$ matrix is unitary, and the missing identity operator (forward scattering amplitude) is missing in the strictly semiclassical treatment). In the same spirit we shall omit the identity operator from the lhs of (65).

Multiplying the semiclassical expression for $S^{(L)}$ by $S^{(R)}$ we get the semiclassical expression for $S$. It coincides with the semiclassical expression for the evolution operator corresponding to the mapping of the transverse section at $x = L$ on itself by trajectories which start at $x = L$ with an initial transverse action $I_i$ and return to $x = L$ after scattering from the boundaries at $x < L$ with transverse action $I_f$. To calculate $S^n$ semiclassically, one has to perform the intermediate sums by the stationary phase method, and one finds that the stationary point condition corresponds to a reflection of the trajectory at the wall $x = L$. That is, the transverse action variable is unaltered upon reflection, while the corresponding angle variable changes its sign. The Trace operation picks up periodic orbits, which were reflected from $x = L$ exactly $n$ times. These are periodic orbits of the original Sinai billiard, but they are chosen according to the multiplicity of their reflections from the transverse line at $x = L$. We finally get the semiclassical expression for the $TrS^n$ as

$$TrS^n \approx \sum_s \frac{p_s}{|\det(I - M_s^r)|^{1/2}} e^{ir(l_s + \pi\nu_s)} \tag{67}$$

Where the sum is over all primitive periodic orbits of the billiard which scatter $p_s$ times from $x = L$ and which satisfy $n = p_s r$. Their length is $l_s$ and $M_s$ is the monodromy matrix. The index $\nu_s$ counts the total number of reflections from the billiard boundaries (including the $n$ reflections from the line $x = L$). This expression completes the semiclassical construction since it can be now substituted in (56, 57, 60) to yield the semiclassical expression for the secular equation.

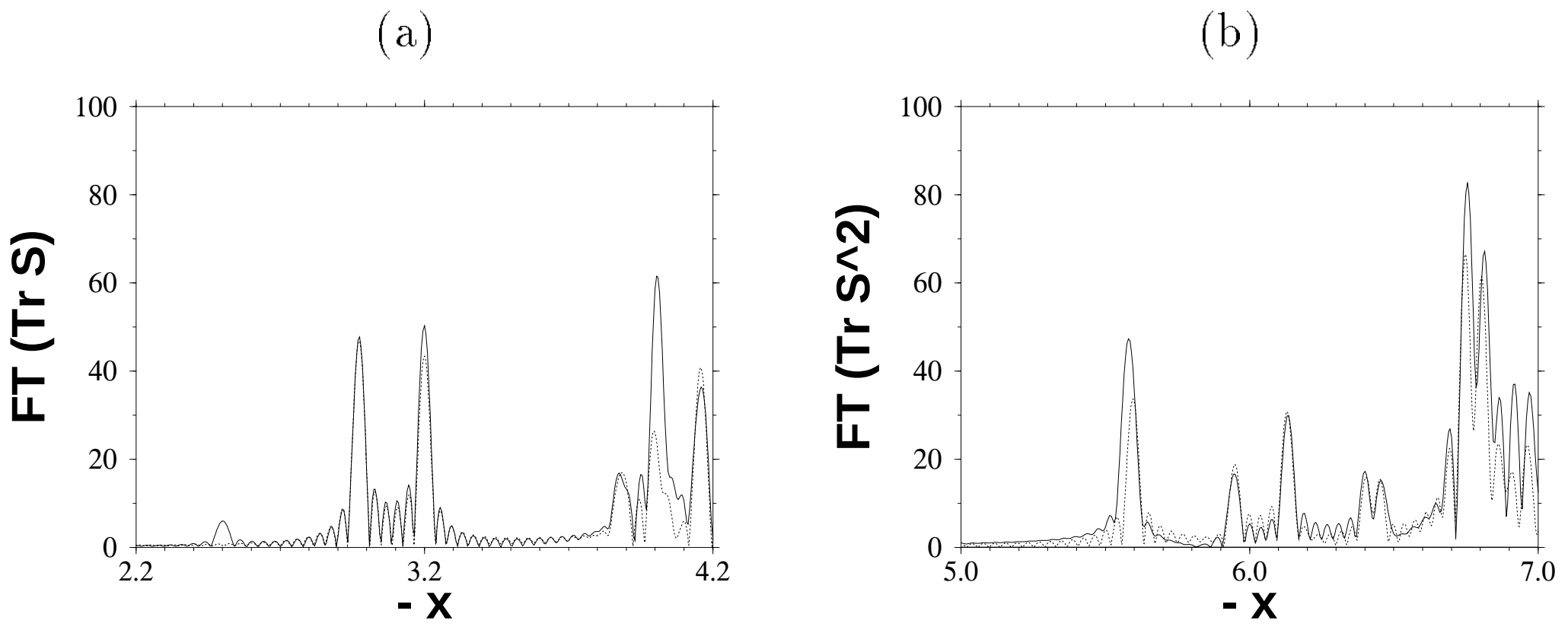

Figure 17: *Fourier transform of* $TrS^N(k)$ *for (a)* $N = 1$ *and (b)* $N = 2$ *(solid line). The shape of the billiard was rendered rectangular (*$L_x = 2$*,* $L_y = 1$*,* $R = 0.75$*) in order to remove degeneracies in the length spectrum. The dashed line shows the semiclassical prediction obtained from some of the shortest unstable PO's.*

# 5 Discussion

The main purpose of this article was to check and demonstrate a few ideas concerning the quantization of billiards using the scattering approach. The system which we considered in detail was the Sinai billiard and we addressed it from various angles. An important point in the present work was the detailed investigation of the rôle of evanescent modes. In this context we would like to recall that the Sinai billiard was previously investigated in [5], where it was quantized by a similar approach, but for one difference, namely, that the channel was not connected to the line $x = L$ but to the section $[R, L]$ of the line $x = 0$ (see figure (1)). It is gratifying that the properties of the system are independent of the method of attaching the auxiliary continuum (the channel). We would like to emphasize that the two scattering problems are very different in character. The attachment of the channel to the narrow section (at $x = 0$) leads to a scattering system which is chaotic, since the set of trapped orbits is a non trivial fractal set. This is not the case when the channel is connected to the right (at $x = L$). The only trapped orbit in this case is the one which reflects between the circle and the straight wall of the billiard along the line $x = 0$. Hence, the scattering problem is not chaotic. Another important difference was explained when the effect of evanescent modes was discussed. It was shown that the increase of the radius has opposite effects in the two quantization schemes. In the present scheme one had to increase the number of evanescent modes as the radius was increased. In the method of [5], the effect of the evanescent modes was diminished.

The scattering approach developed in this article is related in many ways to the method proposed by Bogomolny [8]. As a matter of fact, Bogomolny's approach for billiard problems results in exactly the same semiclassical theory as the one obtained here. The main difference is that here we made an explicit connection with the scattering problem, thus giving the $T$ operator a physical interpretation. The discussion at the end of the previous chapter on the unitarity of the semiclassical $S$ matrix, is relevant to Bogomolny's approach, since also in his case, the $T$ matrix is only semiclassically unitarity. The connection with a scattering problem sheds light on this phenomenon.

The scattering approach to quantization, since it starts from an exact expression can serve as a starting point for deriving improved semiclassical quantization procedures. The first issue to be addressed is the inclusion of the leading corrections which are coming from evanescent modes. Even though these corrections are exponential in $\hbar$, their effect appear in practical calculations as can be seen in figures (13, 14, 15). Other semiclassical corrections, especially those which were recently discussed by Gaspard [20] can be easily incorporated within this approach.

Finally we would like to remark that the same method of quantization which was developed here can be easily extended to two dimensional Hamiltonian problems of the type

$$H(\vec{q},\vec{p}) = \frac{1}{2}p^2 + V(\vec{q}) \tag{68}$$

where $V(\vec{q})$ is a binding potential so that $V(\vec{q}) \to \infty$ for $q \to \infty$. This is done by defining an auxiliary scattering problem in the following way. Assume that the potential $V(\vec{q})$ assumes its minimum value at $\vec{q} = 0$. Construct a scattering problem defined by the potential $\hat{V}(\vec{q}) = V(\vec{q})$ for $q_1 < 0$, and $\hat{V}(\vec{q}) = V(q_1 = 0, q_2)$ for $q_1 \geq 0$. Thus, a channel is obtained, and a scattering matrix $S^L$ can be calculated. A corresponding $S^R$ is built by using the same construction as above for the right part of the potential $V(\vec{q})$. Their product defines the $S$ matrix for the secular equation $det(I - S(E)) = 0$. A detailed discussion of this construct is presented elsewhere.

# Acknowledgements

The research reported in this work was supported by grants from the US Israel Binational Science Foundation, and the Israeli Academy of Sciences. HS acknowledges support from the DFG, the Minerva Foundation and the Minerva Center for Nonlinear Physics. The help and good advice of Mr Harel Primack as well as discussions with Drs. E. Doron, M. Sieber and B. Esser are gratefully acknowledged.

# A Angular momentum decomposition for exponentials

Suppose $k_x^2 + k_y^2 = k^2 > 0$, but $k_x$ and $k_y$ *not necessarily real.* We are looking for the radial functions $R_l(r)$ in the angular momentum decomposition of 2d exponential functions

$$\exp(ik_x x + ik_y y) = \sum_{l=-\infty}^{\infty} R_l(r) e^{il\phi} \tag{69}$$

with $r = \sqrt{x^2+y^2}$, $\cos\phi = x/r$ and $\sin\phi = y/r$. The radial function $R_l(r)$ is a solution of Bessels differential equation and therefore a linear combination of $J_l(kr)$ and $Y_l(kr)$ [11]-(9.1).

$$R_l(r) = r_l J_l(kr) + s_l Y_l(kr) \tag{70}$$

Moreover, it is regular in the origin and thus contains no contribution from the divergent Neumann function: $s_l = 0$. The remaining coefficient $r_l$ will be obtained from the asymptotic behaviour for $r \to 0$. We have in this region [11]

$$J_l(kr) \to \frac{(kr)^l}{2^l l!} \qquad\qquad (l \geq 0) \tag{71}$$

and

$$J_{-l}(kr) = (-1)^l J_l(kr) \tag{72}$$

On the other hand, the Taylor expansion of the exponential reads

$$\begin{aligned} \exp(ik_x x + ik_y y) &= \sum_{l=0}^{\infty} \frac{1}{l!}(ik_x x + ik_y y)^l \\ &= \sum_{l=0}^{\infty} \frac{(ir)^l}{l!}(k_x \cos\phi + k_y \sin\phi)^l \\ &= \sum_{l=0}^{\infty} \frac{(ir)^l}{2^l l!}\Big(k_x(e^{i\phi}+e^{-i\phi}) - ik_y(e^{i\phi}-e^{-i\phi})\Big)^l \\ &= \sum_{l=0}^{\infty} \frac{(ir)^l}{2^l l!}\Big((k_x - ik_y)e^{i\phi} + (k_x+ik_y)e^{-i\phi}\Big)^l \end{aligned} \tag{73}$$

After expanding the sum in powers of $e^{i\phi}$ and neglecting all but the lowest term in r this yields

$$\exp(ik_x x + ik_y y) \to \sum_{l=-\infty}^{\infty} \frac{(ir)^{|l|}}{2^{|l|}|l|!}(k_x - i\,Sign(l) k_y)^{|l|} e^{il\phi}. \tag{74}$$

Comparing this to (71) and (72) we obtain

$$r_l = \left( \frac{i\, Sign(l) k_x + k_y}{k} \right)^{|l|} \tag{75}$$

which was the aim of this appendix. If both, $k_x$ and $k_y$, are real, one can introduce $\phi_k$ through $\cos\phi_k = k_x/k$ and $\sin\phi_k = k_y/k$ and obtains the well known formula

$$\exp(ik_x x + ik_y y) = \sum_{l=-\infty}^{\infty} i^l J_l(kr) e^{il(\phi - \phi_k)} \tag{76}$$

# B Symmetries for the generalized S

In this appendix we show, how unitarity and symmetry of a scattering operator can be completed by similiar relations for the nonunitary parts of the generalized $S$. Moreover we will check the implications of these relations for the numerical calculation of $S^L$ in the system under consideration.

In order to get a compact notation we divide the full $S$ into four parts according to whether the involved indices correspond to open (o) or closed (c) channels:

$$S = \begin{pmatrix} S_{oo} & S_{oc} \\ S_{co} & S_{cc} \end{pmatrix} \tag{77}$$

Because of the real potential, together with (9) it's complex conjugate must be a solution of our problem as well. We assume an evanescent incoming mode $n > \Lambda$ and get

$$\Psi_n^* = \frac{e^{-ik_n x}}{\sqrt{-k_n}} \phi_n - \sum_{m=1}^{\Lambda} S_{mn}^* \frac{e^{-ik_m x}}{\sqrt{k_m}} \phi_m - \sum_{n'=\Lambda+1}^{\infty} S_{n'n}^* \frac{e^{ik_{n'} x}}{\sqrt{-k_{n'}}} \phi_{n'} \tag{78}$$

Any linear combination of $\Psi_n^*$ and $\Psi_n$ is again a solution of our scattering problem. Consider

$$\begin{aligned} i\Psi_n^* - \Psi_n &= i \sum_{m=1}^{\Lambda} S_{mn}^* \frac{e^{-ik_m x}}{\sqrt{k_m}} \phi_m + \\ &\quad + \sum_{m=1}^{\Lambda} S_{mn} \frac{e^{ik_m x}}{\sqrt{k_m}} \phi_m + \sum_{n'=\Lambda+1}^{\infty} (S_{n'n} + S_{n'n}^*) \frac{e^{ik_{n'} x}}{\sqrt{k_{n'}}} \phi_{n'} \end{aligned} \tag{79}$$

As for any solution, $S$ will transform the coefficients of the incoming waves into those of the outgoing ones. Applying this to the last equation we obtain two relations between the different parts of the full $S$:

$$S_{oc} = i\, S_{oo} S_{oc}^* \tag{80}$$

and

$$2\,\Im(S_{cc}) = S_{co}S^*_{oc} \tag{81}$$

Doing the same for an open incoming mode one obtains the well known relation

$$I = S_{oo}S^*_{oo} \tag{82}$$

A similiar idea [19] can be used to generalize the unitarity of $S$. Here we start from (33) and use a rectangular area $\mathcal{G}$ enclosed by the channel walls at $x = 0$, $y = 0$ and $y = L$ together with the line $x = x_0$. If now $\Psi_1 = \Psi_n$ and $\Psi_2 = \Psi_{n'}$ are two solutions in the form (9) only the line $x = x_0$ can contribute to the integral. Moreover, when integrated over $y$ the double sum of modes will collapse into a single sum and we obtain

$$\begin{aligned} 0 = &+ \frac{(e^{-ik_nx})^*}{\sqrt{k_n^*}} \frac{\partial}{\partial x}\left(\delta_{nn'}\frac{e^{-ik_nx}}{\sqrt{k_n}} - S_{nn'}\frac{e^{ik_nx}}{\sqrt{k_n}}\right) - \\ &- \sum_{m=1}^{\infty} S^*_{mn}\frac{(e^{ik_mx})^*}{\sqrt{k_m^*}} \frac{\partial}{\partial x}\left(\delta_{mn'}\frac{e^{-ik_mx}}{\sqrt{k_m}} - S_{mn'}\frac{e^{ik_mx}}{\sqrt{k_m}}\right) - \\ &- \frac{e^{-ik_{n'}x}}{\sqrt{k_{n'}}} \frac{\partial}{\partial x}\left(\delta_{n'n}\frac{(e^{-ik_{n'}x})^*}{\sqrt{k_{n'}^*}} - S^*_{n'n}\frac{(e^{ik_{n'}x})^*}{\sqrt{k_{n'}^*}}\right) + \\ &+ \sum_{m=1}^{\infty} S_{mn'}\frac{e^{ik_mx}}{\sqrt{k_m}} \frac{\partial}{\partial x}\left(\delta_{mn}\frac{(e^{-ik_mx})^*}{\sqrt{k_m^*}} - S^*_{mn}\frac{(e^{ik_mx})^*}{\sqrt{k_m^*}}\right) \end{aligned} \tag{83}$$

There are three different ways to choose $n$ and $n'$: both open, one open and both closed. Each will give an additional relation for $S$ after the evaluation of the complex conjugate in the last equation. They read as follows:

- n, n' open:

$$I = S^T_{oo}S^*_{oo} \tag{84}$$

Together with (82) this implies symmetry and unitarity for the submatrix of $S$ corresponding to open channels.

- n open, n' closed:

$$S^T_{co} = i\,S_{oo}S^*_{oc} \tag{85}$$

Comparing this relation to the previously found (80) we obtain the symmetry for the mixed part of $S$:

$$S_{oc} = S^T_{co} \tag{86}$$

- n, n' closed:

$$S^T_{cc} - S^*_{cc} = i\,S^T_{oc}S^*_{oc} \tag{87}$$

From this we derive using (86) and (81)

$$S_{cc} = S_{cc}^T \tag{88}$$

Thus we have established the symmetry of the full $S$

$$S = S^T \tag{89}$$

and can use it and the additional relations (80) - (82) for testing the numerical accuracy of the calculations. Such a check is particularly important if one deals with evanescent modes since the exponential divergence of the involved numbers might otherwise lead to large numerical errors. However, the obtained relations do not provide a complete check for the generalized $S$ - matrix. To make this point clear we consider the restrictions following from (80) - (82) for $S$ in (42). For this we divide the matrix $C$ in a fashion similiar to (77) into two parts

$$C = \begin{pmatrix} C_o \\ C_c \end{pmatrix} \tag{90}$$

and keep in mind that from the definition (38) we have

$$C_o^* = C_o \tag{91}$$

and

$$C_c^* = iC_c. \tag{92}$$

Now we find from unitarity (82)

$$I \;=\; I \;+\; C_o \left( P^{-1} C_o^T\, C_o (P^{-1})^* \;+\; P^{-1} \;+\; (P^{-1})^* \right) C_o^T \tag{93}$$

$$0 \;=\; P^{-1} C_o^T\, C_o (P^{-1})^* \;+\; P^{-1} \;+\; (P^{-1})^* \tag{94}$$

$$0 \;=\; C_o^T\, C_o \;+\; P^* \;+\; P \tag{95}$$

$$\Re(P) \;=\; -\frac{1}{2} C_o^T\, C_o \tag{96}$$

Similiarly the consistency relation for the mixed part (80) yields

$$C_o\, P^{-1}\, C_c^T \;=\; i\, C_o \Big( (P^{-1})^* + P^{-1}\, C_o^T\, C_o\, (P^{-1})^* \Big) (C_c^T)^* \tag{97}$$

$$P^{-1} \;=\; -(P^{-1})^* - P^{-1}\, C_o^T\, C_o\, (P^{-1})^* \tag{98}$$

and from this follows after multiplication by $P$ and $P^*$ once again (96). The consistency for the purely closed part (81) results in

$$-\, i\, C_c\, P^{-1}\, C_c^T - i\, (C_c)^*\, (P^{-1})^*\, (C_c^T)^* \;=\; -i\, C_c\, P^{-1}\, C_o^T C_o\, (P^{-1})^*\, (C_c^T)^* \tag{99}$$

$$P^{-1} \;+\; (P^{-1})^* \;=\; -\, P^{-1}\, C_o^T C_o\, (P^{-1})^* \tag{100}$$

which is also equivalent to (96). We conclude, that the crucial point in the calculation of $S$ is the computation of $\Im(P)$, the accuracy of which cannot be checked by any of the relations. It contains essentially the imaginary part of the structure function and therefore the algorithm described in the appendix of [10] for its calculation will determine the accuracy of our eigenvalues.

# References

[1] O. Bohigas, M. J. Giannonni, C. Schmit: "Spectral Fluctuations of Classically Chaotic Quantum System"
In T.H. Seligman, H. Nishioka (eds.): "Quantum Chaos and Statistical Nuclear Physics" (Springer - Verlag, Berlin, 1986)

[2] Developments in Boundary Element Methods, eds. Bannerjee, Butterfield (London: Applied science publications), 1980

[3] M.V. Berry: Ann. Phys. 131(1981)163 - 216

[4] J. Koringa: Physica 13(1947)392;
W. Kohn, N. Rostoker: Phys. Rev. 94(1954)1111 - 1124

[5] E. Doron, U. Smilansky: Nonlinearity 5(1992)1055 - 1084

[6] C. A. Pillet: private communication

[7] H. J. Stöckmann, J. Stein: Phys. Rev. Lett. 64(1990)2215

[8] E. B. Bogomolny: Nonlinearity 5(1992)805

[9] W. H. Miller: Adv. Che. Phys. 25(1974)69

[10] R. Blümel, U. Smilansky: Physica 36D(1989)111

[11] M. Abramowitz, I.A. Stegun (eds.): "Pocketbook of Mathematical Functions" (Verlag Harri Deutsch, Frankfurt a. M., 1984)

[12] R. Blümel, U. Smilansky: Phys. Rev. Lett. 64(1990)241

[13] F. J. Dyson: J. Math. Phys. 3(1962)140

[14] H. P. Baltes, E. R. Hilf: "Spectra of Finite Systems" (Bibliographisches Institut, Mannheim, 1976)

[15] M. Sieber, U. Smilansky, S. C. Creagh, R. G. Littlejohn: Non - Generic Spectral Statistics in the Quantized Stadium Billiard J. Phys. A, in press (1993)

[16] M. L. Mehta: Random Matrix Theory (Academic Press, 1967)

[17] C. E. Porter: Statistical Theory of Spectra fluctuations (Academic Press, 1965)

[18] M. V. Berry: Proc. Roy. Soc. A400(1985)229

[19] C. Rouvinez (private communication)

[20] P. Gaspard and D. Alonso: Phys. Rev. A, 47(1993)R3468

[21] H. A. Weidenmüller Studies of Many–Channel Scattering Ann. Phys. 28(1964)60